\documentclass[final,12pt]{colt2024} 

\title[Analyzing the Differentially Private Theil-Sen Estimator]{Analyzing the Differentially Private Theil-Sen Estimator for Simple Linear Regression}
\usepackage{times}

\coltauthor{%
 \Name{Jayshree Sarathy} \Email{js6514@columbia.edu}\\
 \addr Columbia University
 \AND
 \Name{Salil Vadhan} \Email{salil\_vadhan@harvard.edu}\\
 \addr Harvard University
}

\usepackage[utf8]{inputenc} 
\usepackage[T1]{fontenc}    
\usepackage{hyperref}       
\usepackage{booktabs}       
\usepackage{nicefrac}       
\usepackage{microtype}      
\usepackage{makecell}
\usepackage{subcaption}
\usepackage{float}
\usepackage{bbm}
\usepackage{xcolor}
\usepackage[section]{algorithm}
\usepackage{tcolorbox}
\tcbset{width=0.9\textwidth,boxrule=0pt,colback=pink,arc=0pt,auto outer arc,left=0pt,right=0pt,boxsep=5pt}
\usepackage{enumitem}

\newif\ifshowcomments
\showcommentstrue
\newcommand{\mynote}[3]{\marginpar{\tiny \sf \color{#1} {#2}: {#3}}}

\ifshowcomments
\newcommand{\JS}[1]{$\ll$\textsf{\color{purple} \small JS : #1}$\gg$}
\newcommand{\SV}[1]{$\ll$\textsf{\color{green} \small SV : #1}$\gg$}

\newcommand{\removeforgood}[1]{\color{teal}{#1}\color{black}}
\newcommand{\removeforsubmission}[1]{}
\newcommand{\justforsubmission}[1]{#1}
\newcommand{\selfnote}[1]{}

\else 
\renewcommand{\mynote}[3]{}
\newcommand{\JS}[1]{}
\newcommand{\SV}[1]{}

\newcommand{\removeforgood}[1]{}
\newcommand{\removeforsubmission}[1]{}
\newcommand{\justforsubmission}[1]{#1}
\fi
\newcommand{\eqx}[1]{\textcolor{red}{}}

\renewcommand{\selfnote}[1]{}
\newif\ifshowproofs
\showproofsfalse

\newtheorem{mytheorem}{Theorem}[subsection]
\newtheorem{mylemma}[mytheorem]{Lemma}
\newtheorem{fact}[mytheorem]{Fact}

\newtheorem{mycorollary}[mytheorem]{Corollary}

\newtheorem{assumption}[mytheorem]{Assumption}
\newtheorem{mydefinition}[mytheorem]{Definition}

    \newtheorem*{thmn}{Theorem}

\SetKwInput{KwInput}{Input}                
\SetKwInput{KwOutput}{Output}              
\SetKwInput{KwPrivacyparams}{Privacy params}  
\SetKwInput{KwHyperparams}{Hyperparams} 
\SetKwComment{Comment}{/* }{ */}

\newcommand{\pr}{P}
\def\eps{\varepsilon}
\newcommand{\ind}{\mathbbm{1}}

\newcommand{\Var}{\mathrm{Var}}
\newcommand{\Cov}{\mathrm{Cov}}
\newcommand{\E}{\mathrm{E}}

\def\reals{\mathbb{R}}
\def\nonnegreals{\reals_{\geq 0}}
\def\posreals{\reals_{> 0}}
\def\naturals{\mathbb{N}}

\def\calN{\mathcal{N}}
\def\calH{\mathcal{H}}

\newcommand{\bx}{\mathbf{x}}

\def\bx{\textbf{x}}
\def\by{\textbf{y}}
\def\bs{\textbf{s}}

\def\bd{\textit{\textbf{d}}}

\newcommand{\intercept}{\beta_0}
\newcommand{\slope}{\beta_1}

\newcommand{\slopes}{\bs}

\newcommand{\vbin}[1]{\vec{B_{#1}}}
\newcommand{\vbinprime}[1]{\vec{B'_{#1}}}

\def\slopesd{\slopes_{\bd}}

\def\wexpm{\texttt{DPWide}}
\def\wexpmci{\textrm{DPWideCI}}

\def\ts{\textrm{TS}}

\def\tshalf{\textrm{TSHalf}}
\def\dptshalf{\textrm{DPTSHalf}}

\newcommand\qtildewem{\widetilde{q}^{\wexpm}}
\newcommand\qtildewemstar{\widetilde{q}^{\wexpm *}}
\newcommand\betahat{\hat{\slope}}

\newcommand\betahatts{\betahat^{\ts}}
\newcommand\betahattshalf{\betahat^{\tshalf}}

\newcommand\betatildets{\widetilde{\slope}^{\ts}}

\newcommand\betatildewemts{\widetilde{\slope}^{\wexpm\ts}}
\newcommand\betatildetshalf{\widetilde{\slope}^{\dptshalf}}

\newcommand\inv[1]{#1\raisebox{1.15ex}{$\scriptscriptstyle-\!1$}}

\newcommand{\fs}{F_{\slopes}}
\newcommand{\fsinv}{\inv{\fs}}
\newcommand\dist{\textit{dist}}
\newcommand\distms{\dist_{\textrm{ms}}}
\newcommand\distvec{\dist_{\textrm{vec}}}
\newcommand\distpair{\dist_{\textrm{pair-vec}}}

\newcommand\multisets{\textrm{Multisets}}

\newcommand\datauniv{\mathcal{D}}
\newcommand\dataunivreal{\reals \times \reals}
\newcommand\datasize{n}
\newcommand\dataspace{\multisets(\datauniv, \datasize)}
\newcommand\dataspacereal{\multisets(\dataunivreal, \datasize)}
\newcommand\datatuples{(x_i, y_i)_{i=1}^n}
\newcommand\binonesize{m_1}
\newcommand\bintwosize{m_2}
\newcommand\binspace{(\datauniv^{\binonesize}, \datauniv^{\bintwosize})}
\newcommand\halfpairs{\lfloor n/2 \rfloor}
\newcommand\halfpairsceil{\lceil n/2 \rceil}
\newcommand\binonesizereal{\halfpairs}
\newcommand\bintwosizereal{\halfpairsceil}
\newcommand\binspacereal{( (\dataunivreal)^{\binonesizereal}, (\dataunivreal)^{\bintwosizereal})}

\newcommand{\privCIL}[1]{\widetilde{L_{#1}}} 
\newcommand{\privCIU}[1]{\widetilde{U_{#1}}} 

\newcommand{\range}{R}
\newcommand{\rangeinput}{[-\range, \range] \subset \reals}

\newcommand{\alphainput}{\alpha \in (0,1)}
\newcommand{\epsinput}{\eps \in \posreals}
\newcommand{\graninput}{\theta \in \posreals}

\newcommand{\nonprivconv}[1]{c_{#1}}

\newcommand{\guess}{\delta}
\newcommand{\guessn}{\guess_n}

\newcommand{\numpairs}{N}
\newcommand{\numpairsdenom}{\numpairs}

\newcommand{\matchedindexset}{S}
\newcommand{\pairsumdouble}{\{i, j\} \in \matchedindexset}

\newcommand{\pairsum}{\{i, j\} \in \matchedindexset, i < j}

\newcommand{\pairsumst}{\{s, t\} \in \matchedindexset, s < t}

\newcommand{\pairsumoverlap}[1]{|\{i, j\} \cap \{s, t\}| = #1}

\newcommand{\sign}{\text{sign}}

\newcommand{\es}{e_1, \ldots, e_n}
\newcommand{\fdiff}{F_{\textrm{diff}}}

\newcommand{\denommu}{\mu_\Psi}
\newcommand{\denomvar}{\sigma_\Psi}

\def\ncov{n\text{cov}}
\def\nvar{n\text{var}}

\newcommand{\splitdesign}{\text{asymptotically optimal design}}

\allowdisplaybreaks

\begin{document}

\maketitle

\begin{abstract}%
In this paper, we study differentially private point and confidence interval estimators for simple linear regression. Motivated by recent work that highlights the strong empirical performance of an algorithm based on robust statistics, \texttt{DPTheilSen}, we provide a  rigorous, finite-sample analysis of its privacy and accuracy properties, offer guidance on setting hyperparameters, 
and show how to produce  differentially private confidence intervals to accompany its point estimates.
\end{abstract}

\begin{keywords}%
  List of keywords%
\end{keywords}

\section{Introduction}
\removeforsubmission{Science and social science research often requires analyzing sensitive datasets. However, decades of study and real-world attacks have shown that release of statistical information -- if too many and too accurate -- can allow for reconstruction of the data subjects' sensitive attributes~\cite{DN03,netflix1}.} 
\removeforsubmission{One promising solution is \emph{differential privacy} (DP)~\cite{Dwork06}, a rigorous mathematical framework for characterizing privacy loss.}
\justforsubmission{Over the last several years, differential privacy (DP)~\citep{Dwork06} has become a widely accepted standard for protecting the privacy of data subjects while releasing useful statistical information about datasets~\citep{abowd2018us,altman2018practical,biden2023executive}. However, designing \emph{usable} DP methods for common statistical inference tasks remains an ongoing challenge~\citep{barrientos2021feasibility,sarathy2023don,garrido2023lessons}.}~One example is simple (ie. one-dimensional) linear regression, which is one of the most fundamental tasks in data analysis. 
In 2018, the prominent economics research group, Opportunity Insights, found that there was a lack of guidance around choosing accurate DP algorithms for simple linear regression on regimes commonly used in practice (e.g. small-area analysis with $40$ to $400$ datapoints per regression)~\citep{ChettyF19}. 

Motivated by this gap between theory and practice,~\cite{alabi2022differentially} conducted an empirical evaluation of several DP algorithms for simple linear regression. They found that a suite of robust, median-based algorithms, \texttt{DPTheilSen}, based on the non-private Theil-Sen estimator developed by~\cite{Theil50} and~\cite{Sen68}, performed better than standard OLS-based algorithms across a range of practical regimes. 
\texttt{DPTheilSen} operates in two steps: first, it computes the slope for some or all pairs of points and second, uses a DP median algorithm to output a single estimate of the slope.\footnote{The intercept can be computed either along with the slope or using the final estimate of the slope, as described in~\citep{Sen68} and~\citep{alabi2022differentially}.}

While Alabi et al.'s empirical study was a valuable starting point, the authors stated that further theoretical understanding of the accuracy guarantees of \texttt{DPTheilSen}, as well as design of uncertainty estimates, would be needed to make this set of algorithms fully usable in practice.
In this paper, we address these open questions by analyzing the accuracy guarantees of the \texttt{DPTheilSen} algorithms.
Our contributions can be summarized as follows.
\begin{enumerate}
    \item We provide a rigorous theoretical analysis of the \texttt{DPTheilSen} algorithms shown to perform strongly by~\cite{alabi2022differentially}. Our analyses offer finite-sample convergence bounds and shed light on \emph{why} and \emph{when} \texttt{DPTheilSen} outperforms other DP linear regression algorithms. 
    \item We design and analyze DP confidence intervals for simple linear regression via \texttt{DPTheilSen}. 
\end{enumerate}

\removeforsubmission{We note that~\cite{DworkL09} analyzed one version of \texttt{DPTheilSen} in the privacy setting, which they called the ``Short-Cut Regression Method," but they did not analyze the variants that\removeforsubmission{ were shown by~\cite{alabi2022differentially} to} exhibit stronger performance. They provide an asymptotic rather than a finite-sample analysis, and they did not consider uncertainty measures such as confidence intervals.}

Although our work focuses on the one-dimensional linear regression setting, our approach is a crucial starting point to developing similar finite-sample analyses for higher-dimensional regimes.\footnote{In particular, our analysis can be used as a black box to analyze~\cite{knop2022differentially}'s approach of coordinate-wise medians for higher-dimensional linear regression. Some new ideas may be required to analyze approaches that use more sophisticated higher-dimensional aggregators based on Tukey depth (see Section~\ref{sec:related-work}), but again our analysis should provide guidance as it is the one-dimensional analogue of those methods.} 
Overall, we view our work as a contribution towards understanding private and robust regression more generally.
Before providing an overview of our main results, we describe the problem of simple linear regression and the assumptions we make in this work, which are minimal.

\subsection{Simple linear regression}
\label{sec:linreg}
We are given $n$ values, $x_1, \ldots, x_n$, of the predictor variable $x$. For each 
$x_i$, we observe the corresponding value $y_i$ of the response random variable $y$. The model is $y_i = \intercept + \slope x_i + e_i$ for $i = 1, \ldots, n$, where $\intercept$  and $\slope$ are unknown parameters.
We make the following assumptions:
\begin{assumption}
\label{assumption:x-constants}
    $x_1, \ldots, x_n$ are fixed and not all equal.
\end{assumption}
\begin{assumption}
\label{assumption:noise-variables}
    Each $e_i$ is sampled independently from the same continuous,
    symmetric, mean-0 distribution $F_e$. 
\end{assumption}

\begin{figure}[hbtp]
\centering
\includegraphics[width=0.35\textwidth]{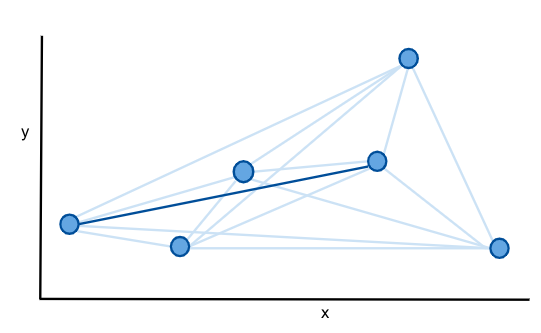}
\caption{Illustration of the standard non-private Theil-Sen algorithm~\citep{Theil50,Sen68}, which (1) computes the slopes between all pairs of points, and (2) outputs the median slope in dark blue.}
\label{fig:ts}
\end{figure}

A common formulation of linear regression is the Ordinary Least Squares (OLS) objective, which is characterized by the following optimization problem:
\begin{equation*}\label{eq:OLSopt}
(\hat\intercept, \hat\slope)=\arg\min_{\intercept,\slope\in\reals} \|\by-\slope\bx-\intercept\textbf{1}\|_2,
\end{equation*}
where $\bx = (x_1, \ldots, x_n)^T$, $\by = (y_1, \ldots, y_n)^T$, and $\textbf{1}$ is the all-ones
vector. OLS has a simple closed form solution:
\begin{equation*}\label{eq:OLSclosedform}
\hat\slope = \frac{\ncov(\bx, \by)}{\nvar(\bx)}\;\; \text{and} \;\; \hat\intercept = \bar{y} - \hat\slope\bar{x},
\end{equation*}
where $\bar{x} = \frac{1}{n}\sum_{i=1}^n x_i$,  $\bar{y} = \frac{1}{n}\sum_{i=1}^n y_i$, $\ncov(\bx, \by) = \langle \bx-\bar{x}\mathbf{1}, \by-\bar{y}\mathbf{1}\rangle,$ and $\nvar(\bx) = \langle \bx-\bar{x}\mathbf{1}, \bx-\bar{x}\mathbf{1}\rangle.$
When $\by$ is
generated according to the model $y_i = \intercept + \slope\cdot x_i + e_i, \forall i\in[n]$ for i.i.d.
Gaussian noise $e_i$, then the OLS solution is the maximum likelihood estimator. If we remove the assumption of Gaussian noise and add privacy constraints, however, robust estimators such as Theil-Sen have been shown to provide better accuracy~\citep{DworkL09,alabi2022differentially}. 
In this work, we analyze DP algorithms based on Theil-Sen, comparing them to the OLS-based approaches. 
Below, we provide brief descriptions of these two approaches:
\begin{itemize}
    \item \textbf{OLS-based algorithms}: The \texttt{DPSuffStats} algorithm~\citep{DworkTT014,Wang18,alabi2022differentially} follows the OLS approach closely, in that it involves perturbing the sufficient statistics $\ncov(\bx,\by)$ and $\nvar(\bx)$. While this algorithm is computationally efficient and enables releasing the DP \removeforsubmission{versions of the }sufficient statistics at no extra privacy cost, it has been shown to not perform well in common regimes.\footnote{In particular,~\cite{alabi2022differentially} showed that \texttt{DPSuffStats} performs poorly in the high privacy, small dataset, and/or clustered independent variable regime.}A second algorithm is \texttt{DPOLSExp}~\citep{bassily2014private,alabi2022algorithmic}, which implements the exponential mechanism~\citep{McSherryT07} with the OLS objective function. 
    \item \textbf{Theil-Sen-based algorithms}: Theil-Sen~\citep{Theil50,Sen68} is a family of robust linear regression estimators that proceed via two steps (as illustrated in Fig.~\ref{fig:ts}) for estimating the slope $\slope$: (1) compute the slope for some or all pairs of points, and (2) output the median of these estimates. To make this algorithm DP, we can simply replace the median with a DP median.
    Although there are many choices for the DP median, we consider
    a version of this algorithm, \texttt{DPWideTS}, which uses the \emph{widened exponential mechanism}~\citep{alabi2022differentially,McSherryT07} as the DP median sub-routine, as this version has been shown to exhibit strong empirical performance by~\cite{alabi2022differentially}.
    \removeforsubmission{To produce DP confidence intervals, we run \texttt{DPWideTS} twice such that with high probability, the two estimates capture the true slope $\slope$.}
\end{itemize} 

\subsection{Overview of results}
 
This work offers two main contributions: finite-sample guarantees and confidence intervals for \texttt{DPTheilSen}.
First, we provide a finite-sample convergence bound for \texttt{DPWideTS}, the variant of DP Theil-Sen that was recommended by~\cite{alabi2022differentially}. We attain the convergence bound by developing a finite-sample analysis of non-i.i.d. \emph{U-statistics}~\citep{Hoe48,chen2011normal}, which may be of independent interest.

We state our convergence bound informally below. While our main analysis does not require any data assumptions beyond the ones stated above (Assumptions~\ref{assumption:x-constants} and~\ref{assumption:noise-variables}), the theorem is stated for a special case, where the independent variables $x_1, \ldots, x_n$ are evenly split between the endpoints of an interval of length $\Delta_x$ (known as an \emph{\splitdesign}~\citep{Sen68}), and the noise variables $e_1, \ldots, e_n$ are drawn i.i.d. from $\calN(0, \sigma_e^2)$. 

\begin{mytheorem}[Main result applied for case of \splitdesign, informally stated]
\label{thm:wemts-bound-endpoints}
Let $\betatildewemts$ be the DPWideTS estimator with privacy loss parameter $\eps$, hyperparameter $R$ for the range of the outputs, and hyperparameter $\theta$ for the granularity of the outputs. Assume that the true slope $\slope$ lies in the interval $[-R+\theta, R-\theta]$.
Let $\tau$ be defined as follows.
\begin{align*}
    \tau = \Phi^{-1} \left( 1 - \frac{p}{8} \right) \cdot \sqrt{\frac{4}{3n}} + O \left( \frac{\ln (R/p\theta)}{\eps n} \right)
\end{align*}
where $\Phi^{-1}$ is the inverse standard normal distribution function.
Then, for suff. large $n$, and suff. small $\tau$ and $\sigma_e/\Delta_x$,  we have that with probability at least $1-p$, 
\begin{align*}
\betatildewemts \in [\slope - z - \theta, \slope + z + \theta] \\
    \text{for}~ z = \frac{\sqrt{\pi} \cdot \sigma_e}{\Delta_x} \cdot \left( \tau + O\left(\tau^{3/2}\right) \right).
\end{align*}
\end{mytheorem}
In Table~\ref{tab:comparison-intro} below, we provide an informal comparison between the convergence bounds for \texttt{DPWideTS} (described above) and the OLS-based \texttt{DPSuffStats} algorithm for this setting (analyzed in~\cite{alabi2022algorithmic}). We also compare these DP estimators with their non-private counterparts. \removeforsubmission{Notation is provided in the caption.}We can see that \texttt{DPWideTS} maintains the same leading constant of $2 \sqrt{\pi/3}$ as that of non-private TS, while non-private OLS and \texttt{DPSuffStats} have a leading constant of $2$, so the latter are slightly better in the asymptotic $n \rightarrow \infty$ regime. 
For \texttt{DPWideTS}, we see a logarithmic dependence on the output range $R$, as compared to the quadratic dependence shown by \texttt{DPSuffStats} on the input range $r$. Our comparison provides theoretical backing to the empirical finding of~\cite{alabi2022differentially}---that the quantity $\eps n \Delta_x^2$ is important in choosing between the standard \texttt{DPSuffStats} algorithm and the robust \texttt{DPWideTS} algorithm. In particular, when $\eps n \Delta_x^2$ is small, which indicates a high privacy, small dataset size, and/or clustered independent variable regime, \texttt{DPWideTS} is the more accurate estimator. Finally, our convergence bound enables us to provide guidance on setting the $\theta$ hyperparameter (corresponding to the granularity of the widened exponential mechanism) for $\texttt{DPWideTS}$, which was a key open problem raised by~\cite{alabi2022differentially} towards making this algorithm usable for practitioners. 

\begin{table}[hptb!]
\centering
  \begin{tabular}{ | c | c |  } 
    \hline 
    & \\ [-.2ex]
    \small Estimator & \small $1-p$ Convergence Bound \\ [1ex]
    \hline 
    & \\ [-.2ex]
    \small Non-priv OLS &
    $ \frac{2 \sigma_e}{\Delta_x} \cdot \frac{\nonprivconv{p/2}}{\sqrt{n}} $ \\ [1.5ex]
    \hline
    & \\ [-.2ex]
    \small Non-priv TS & $ \sqrt{\frac{\pi}{3}} \cdot \frac{2\sigma_e}{\Delta_x} \cdot \frac{\nonprivconv{p/4}}{\sqrt{n}} $ \\ [1.5ex]
    \hline
    & \\ [-.2ex]
    \small \texttt{DPSuffStats}  &
    \small $ \frac{2 \sigma_e}{\Delta_x} \cdot \frac{\nonprivconv{p/6}}{\sqrt{n}} \cdot \left( 1 + \tau \right) + \tau \left( 1 + \tau + |\slope| \right)$,
    \\
    \small \citep{alabi2022algorithmic} & \small $\tau \approx \frac{(1-1/n)r^2 \log(3/p)}{\eps \cdot n \cdot \Delta_x^2}$
    \\ [1.5ex]
    \hline
    &  \\ [-.2ex]
    \small \texttt{DPWideTS}  & 
    \small $\sqrt{\frac{\pi}{3}} \cdot \frac{2\sigma_e}{\Delta_x} \cdot \left( \frac{\nonprivconv{p/8}}{\sqrt{n}} + \gamma \right) (1 + o(1))$ \\ 
    \small (Thm.~\ref{thm:wemts-bound-endpoints}) & 
    \small $+~ \theta, ~ \gamma \approx \frac{\ln \left( R / p \cdot \theta \right)}{\eps n}$ \\
     [1.5ex]
    \hline
  \end{tabular} 
  \\ [1.5ex]
\caption{$1-p$ convergence bounds for point estimators for simple linear regression in special case of \splitdesign. Note that $r$ is range for both the input $x_i, y_i$ datapoints, $R$ is range for the output estimate of $\slope$, and $\nonprivconv{p/2} = \Phi^{-1}(1-p/2) = \Theta(\log(1/p)$ for small $p$.}
\label{tab:comparison-intro}
\end{table}
In addition to analyzing the \texttt{DPWideTS} point estimator, we design and analyze corresponding confidence interval estimators for simple linear regression. We describe two algorithms---\texttt{DPWideTSCIUnion} and \texttt{DPWideTSCI}---and analyze their privacy, coverage, and accuracy guarantees. 
We use the DP median estimators proposed by~\cite{DGMSS21} as sub-routines in our algorithms, making them additionally usable by analyzing the width of the confidence intervals they provide. 
We show that the confidence interval for \texttt{DPWideTSCIUnion} is approximately twice as large as the convergence bound for \texttt{DPWideTS}, and can be further improved using the algorithm \texttt{DPWideTSCI}\removeforsubmission{ with some tradeoff in computational efficiency}. 

\subsection{Related work}
\label{sec:related-work}

This work draws on the rich connections between robust statistics and DP~\citep{DworkL09,asi2023robustness,hopkins2023robustness}.~\cite{DworkL09}\removeforsubmission{stated that ``robust estimators are a useful starting point for constructing highly accurate differentially private estimators.'' They} gave an asymptotic analysis of what they called the ``Short-Cut Regression Method," which is similar to a simplified variant of \texttt{DPTheilSen}.
However, they did not consider the more statistically efficient variants 
that we do in this work, provide finite sample guarantees, or offer measures of uncertainty for the estimates like the confidence intervals we provide.
Other works~\citep{CKSBG19,avella2021privacy} confirm the findings of Dwork and Lei in the context of hypothesis testing\justforsubmission{ but similarly do not provide finite-sample analysis. A longer discussion of these works is provided in Appendix~\ref{sec:related-work-cont}.}
\removeforsubmission{, showing that robust estimators perform better than parametric estimators under differential privacy, even when the data comes from a parametric model,} 
\removeforsubmission{, but these works provide either empirical or asymptotic guarantees rather than finite-sample analysis.}

In general, linear regression is one of the most fundamental tasks in statistics and yet has been a challenging problem in the DP literature~\citep{ZhangZXYW12,Sheffet17,Barrientos:2017,Bernstein:2018,AwanS18,Wang18,CWZ19,Evans:2021,ramsay2021differentially,knop2022differentially,liu2022differential,liu2023label}. Most prior works focus on non-robust methods, require additional assumptions on the data or model, or only provide asymptotic analysis.
To address these gaps,~\cite{alabi2022differentially} conducted an experimental evaluation of DP algorithms for simple linear regression in the \emph{small dataset regime}, demonstrating that DP analogues of robust algorithms, such as \texttt{DPTheilSen}, perform better than non-robust methods\justforsubmission{ in practical settings. }\removeforsubmission{ when the dataset size, variance of the independent variables, and/or privacy loss parameter is small.} However, Alabi et al. did not provide theoretical bounds or DP confidence intervals for this estimator, which are open questions we address in this work.

Our confidence interval algorithms build upon~\cite{DGMSS21}'s non-parametric DP confidence intervals for the median, but ours are more general in that they provide finite-sample validity for 
some forms of non-i.i.d. variables 
and characterizes the width of the confidence intervals. 
\removeforsubmission{
Prior work has also considered DP confidence intervals for mean estimation~\cite{Karwa:2018,Gaboardi:2018,Du:2020}, but these cannot directly be applied for the median-based estimator we consider.}
Other work 
produces DP confidence intervals using bootstrapping~\citep{ DOrazio:2015,Barrientos:2017,Brawner:2018,Ferrando:2020}, but these can be expensive to compute and rely on assumptions, such as normally distributed errors, 
that our work avoids. 
\removeforsubmission{
To the best of our knowledge, our work is the first to theoretically analyze the different variants of \texttt{DPTheilSen} as well as to design and analyze non-parametric, differentially private confidence intervals to accompany its point estimates.
}

\textit{Concurrent work.} Since our results were first announced,\footnote{References for presentations of our results in 2021 have been removed for anonymity.} there has been a flurry of work that uses 
median-based approaches for DP tasks~\citep{brown2021covariance,liu2022differential,ramsay2021differentially,cumings2022differentially,knop2022differentially} and clarifies the connections between privacy and robustness~\citep{asi2023robustness,hopkins2023robustness}.\footnote{See Appendix~\ref{sec:related-work-cont} for a longer discussion of related and concurrent work.} 
None provides the analysis we offer in this paper;
we touch on a few here but do not provide a comprehensive survey.

One relevant work by \cite{amin2023easy} provides empirical evaluation of higher-dimensional variations of Theil-Sen, but unlike our work, does not offer theoretical bounds on accuracy or convergence. \cite{knop2022differentially}, on the other hand, do
provide theoretical privacy and utility analysis of some of the DP Theil-Sen algorithms in higher dimensions, as well as an experimental evaluation.  
They offer a finite-sample convergence bound that is consistent with our results, but their statement only applies for the highly simplified variant of \texttt{DPTheilSen} where each data point is used only once, so only $n/2$ slopes are computed and the slopes are independent of each other. Our results are more general and can handle cases that achieve provably stronger performance by reusing data points and introducing correlations. Examples include the asymptotically optimal design of Thm.~\ref{thm:wemts-bound-endpoints} and the full Theil-Sen variant where all ${n \choose 2}$ points are used.\footnote{Note that the simplified variant of Theil-Sen can be analyzed using a Chernoff-Hoeffding bound, while the versions of Theil-Sen with correlated slopes require more complex analytical tools such as U-statistics.} In addition, Knop and Steinke
do not aim to match the leading constant of the non-private estimators, which is an important feature of our results. 

\removeforsubmission{
Recent work has also significantly advanced our understanding of the connection between robust and private algorithms.
Work by Asi, Ullman, and Zakynthinou~\cite{asi2023robustness} establishes a tight connection between privacy and robustness, providing a 
black-box transformation from optimal robust to optimal differentially private algorithms. They also design and analyze estimators for DP linear regression for high-dimensional tasks under assumptions of Gaussianity. 
Work by Hopkins et al.~\cite{hopkins2023robustness} shows how to implement this black-box transformation in a computationally efficient manner via the sum-of-squares method. While these methods provide important general toolkits for DP algorithm design, they do not provide finite-sample bounds for specific estimators such as \texttt{DPTheilSen}.
}

\removeforsubmission{
More generally, recent studies have further highlighted the need for more usable DP algorithms for basic statistical tasks such as simple linear regression.}
\removeforsubmission{
A recent study by Barrientos et al.~\cite{barrientos2021feasibility} evaluates a range of DP linear regression algorithms in terms of feasibility for real-world use. Their criteria includes: assumptions that align with practical applications, ease of implementation, computational efficiency, minimal tuning parameters, and accompanied by uncertainty estimates. Barrientos et al. find that a suprisingly few number of DP algorithms satisfy these criteria. Our work on developing the design and analysis of \texttt{DPTheilSen} makes this suite of algorithms measure up across all of these categories, suggesting that \texttt{DPTheilSen} will now be usable for real-world DP deployments.
}

\section{Preliminaries}
We consider datasets that are multisets. The space of datasets is denoted by $\dataspace$, where $\datauniv$ is the underlying set of elements and $\datasize$ is the cardinality of each multiset.\justforsubmission{ Let $\distms(\bd, \bd')$ be the number of records that must be changed to transform $\bd$ into $\bd'$.\footnote{I.e. ``change-one distance''~\citep{casacuberta2022widespread}).}}
\removeforsubmission{We view datasets $\bd \in \dataspace$ as specified by histograms $m_{\bd}: \datauniv \rightarrow \naturals$
where $m_{\bd}(w)$ gives the multiplicity of element $w$ (so $\sum_{w \in \datauniv} m_{\bd}(w) = \datasize$).
We define the distance function for multisets, $\distms : \dataspace \times \dataspace \rightarrow \naturals$, as follows.\footnote{This is also called ``change-one distance''~\cite{casacuberta2022widespread}.}
\begin{mydefinition}
For any two datasets $\bd, \bd' \in \dataspace$, the \emph{distance function} is
\[\distms(\bd, \bd') = \frac{1}{2} \sum_{w \in \datauniv} |m_{\bd}(w) - m_{\bd'}(w)|
\] 
$\distms(\bd, \bd')$ is the number of records that need to be changed to transform $\bd$ into $\bd'$.
\end{mydefinition}
}

\removeforsubmission{Let us now define the notion of differential privacy (DP) we use.} 
Since our algorithms include hyperparameters, we state a definition of DP for algorithms that take as input not only the dataset, but also the desired privacy parameters and any required hyperparameters. 
\removeforsubmission{Let $\datauniv$ be a data universe and $\dataspace$ be the space of datasets.} Two datasets $\bd, \bd' \in \dataspace$ are \emph{neighboring}, denoted $\bd \sim \bd'$, if 
$\dist(\bd, \bd') = 1$.
Let $\mathcal{H}$ be a hyperparameter space and $\mathcal{Y}$ be an output space.


\begin{mydefinition}[Differential Privacy~\citep{DMNS06}]\label{def:dp-with-inputs}
For $\eps \in \reals_{\geq 0}$, a randomized algorithm $M: \dataspace \times \reals_{\geq 0} \times \mathcal{H} \rightarrow \mathcal{Y}$ is \emph{$\eps$-DP} if and only if for all neighboring datasets $\bd \sim \bd' \in \dataspace$
$\text{hyperparams} \in \mathcal{H}$, and sets $E \subseteq \mathcal{Y}$, 
\begin{align*} 
    \Pr&[M(\bd, \eps, \text{hyperparams}) \in E]  \\
    &\leq e^\eps \cdot \Pr[M(\bd', \eps, \text{hyperparams}) \in E].
\end{align*}
where the probabilities are taken over the random coins of $M$.
\end{mydefinition}

Now, we will define the non-private Theil-Sen family of estimators. 
\begin{mydefinition}[Theil-Sen Estimator~\citep{Theil50,Sen68}]
Let $(x_1, y_1), \ldots, (x_n, y_n)$ be an arbitrary ordering of dataset $\bd \in \dataspacereal$. 
Let $\matchedindexset$ be a set of $\numpairs$ unordered pairs of elements of $[n] = \{1, \ldots, n\}$ such that for each pair $\{i, j\} \in \matchedindexset$, $x_i \neq x_j$.
Then, for each $\{i, j\} \in \matchedindexset$,
compute the slope $s_{ij}$ between the points $(x_i, y_i)$ and $(x_j, y_j)$ as follows:
 $s_{ij} = (y_j - y_i)/(x_j - x_i)$.
Let $\slopes = \{s_{ij}\}_{\{i, j\} \in \matchedindexset}$ denote the multiset of the slopes. The \emph{Theil-Sen estimator $\betahatts$} with corresponding set $\matchedindexset$ is computed as follows: $\betahatts = \textrm{median}(\slopes)$.
\end{mydefinition}

\section{\texttt{DPTheilSen} Algorithm}
In the differentially private version of Theil-Sen, which we call \texttt{DPTheilSen} (Algorithm~\ref{alg:ts}), we similarly compute pairwise estimates of the slope. However, we replacing the computation of the median of the slopes with a differentially private median algorithm (denoted by DPmed, which can be one of several algorithms). DPmed takes as input the multiset of slopes, $\slopesd \in \multisets(\reals \cup \{-\infty, \infty\}, \numpairs)$, the scaled privacy parameter $\eps / k \in \nonnegreals$, where $k$ is the max number of slopes computed using each datapoint, and the hyperparameters $\in \calH$ for the given median algorithm. 





\begin{algorithm}[h!]
  \KwData{$\bd = \datatuples \in \dataspacereal$, }
  \KwPrivacyparams{$\eps \in \nonnegreals$}
  \KwHyperparams{$\matchedindexset \in \binom{[n]}{2}$, DPmed, $\text{hyperparams} \in \calH$}

  $\slopes = \{\}$ 

  \For{each $\{i,j\} \in \matchedindexset$ (such that $x_i \neq x_j$)}{
    

    $s_{ij} = (y_j - y_i)/(x_j - x_i)$
    
    Add $s_{ij}$ to $\slopes$

  }
  Let $k = \max_{i \in [n]} \{ \# j \in [n]: \{i, j\} \in \matchedindexset \}$
  
  $\betatildets = \textrm{DPmed}\left(\slopes, \eps/k, \text{hyperparams}\right)$
  
  \Return $\betatildets$
  \caption{\texttt{DPTheilSen}: $\eps$-DP Algorithm}
  \label{alg:ts}
\end{algorithm}

\begin{mylemma}[\cite{alabi2022differentially}]
    Algorithm~\ref{alg:ts} (\texttt{DPTheilSen}) is $\eps$-DP.
\end{mylemma}
In this work, we consider a version of \texttt{DPTheilSen} called \texttt{DPWideTS}, which uses for DPmed the widened exponential mechanism, \texttt{DPWide}~\citep{alabi2022differentially,McSherryT07}.
This variant of \texttt{DPTheilSen} was found to have strong performance in the empirical work of~\cite{alabi2022differentially}.
\justforsubmission{The \texttt{DPWide} algorithm (Alg.~\ref{alg:wem}) and utility theorem (Thm.~\ref{thm:utility-of-wem}) are included in the Appendix.}

\removeforsubmission{
\begin{algorithm}[h!]
  \KwData{$\slopes = (s_1, \ldots, s_\numpairs) \in \reals^\numpairs$}
  
  \KwPrivacyparams{$\eps \in \nonnegreals$}
  
  \KwHyperparams{$q \in (0,1), -R, R \in \reals^2; \graninput$}
  
  Sort $\slopes$
  
  \Comment{Clip $\slopes$ to the range $[-R, R]$ and insert space $\theta$ around the $q$th quantile:}
    

    
  
  \For{$i \in [1, \lfloor \numpairs q \rfloor ]$}{
    $\slopes[i] = \min(\max(-R, \slopes[i] - \theta), R)$
   }
   
   \For{$i \in [\lfloor \numpairs q \rfloor + 1, \numpairs]$}{
    
    $\slopes[i] = \max(\min(R, \slopes[i] + \theta), -R)$
  }
  
  Insert $-R$ and $R$ into $\slopes$ and set $\numpairs = \numpairs+2$
  
  Set $\textrm{maxNoisyScore} = -\infty$
  
  Set $\textrm{argMaxNoisyScore} = -1$

  \For{$i \in [2, \numpairs)$} {
    
    
    $\textrm{score} = \log(\slopes[i] - \slopes[i-1]) -\frac{\eps}{2} \cdot \lfloor | i - \numpairs q |\rfloor $
    
    $Z \sim \textrm{Gumbel}(0,1)$ 
    
    $\textrm{noisyScore} = \textrm{score} + Z$
    
    \If {$\textrm{noisyScore} > \textrm{maxNoisyScore}$} {
        $\textrm{maxNoisyScore}= \textrm{noisyScore}$
        
        $\text{argMaxNoisyScore} = i$
        }
  }
  
  $\text{left} = \slopes[\text{argMaxNoisyScore}-1]$
  
  $\text{right} = \slopes[\text{argMaxNoisyScore}]$
  
  Sample $\qtildewem \sim \text{Unif}\left[\text{left} , \text{right} \right]$
  
  \Return $\qtildewem$

    \caption{Widened Exponential Mechanism for Quantile ($\wexpm$): $\eps$-DP Algorithm}  \label{alg:wem}
\end{algorithm}

Below, we state the standard utility theorem for $\wexpm$ with widening hyperparameter $\theta$ on a fixed dataset $\slopes$. Let $\fsinv$ be the inverse empirical distribution function for the set of slopes, $\slopes$.  
We use the following assumption on the range and widening hyperparameters $R, \theta \in \posreals$. 
\begin{assumption} 
\label{assumption:range}
For a target $q \in (0,1)$ and fixed dataset $\slopes \in \reals^N$, the true quantile $\fsinv(q) \in [- R + \theta, R - \theta]$.
\end{assumption}

\begin{theorem}[Utility of \wexpm]
\label{thm:utility-of-wem}
Let $\slopes \in \reals^\numpairs$ be a fixed sample of $\numpairs$ points, and let $\fsinv$ be the inverse empirical distribution function. For $q \in (0,1)$, let $[-R,R]$, the range hyperparameter, satisfy Assumption~\ref{assumption:range}, and let $\theta, \eps > 0$. Let $\qtildewem = \texttt{DPWide}(\slopes, \eps, (q, -R, R, \theta)$.
Then, for $0 < c < \min(q, 1/2)$, 
\begin{align*}
\Pr_{ \substack{\wexpm(\slopes, \eps, \\ (q, -R, R, \theta))}}
\Bigl( &\qtildewem \notin [\fsinv(q - c) - \theta, \fsinv(q + c) + \theta] \Bigr) \\
&\leq \frac{R}{\theta} \exp \left(-\frac{\varepsilon c\numpairs}{2}\right)
\end{align*}
\end{theorem}
}

\section{Convergence Bound for \texttt{DPWideTS}}
In this section, we provide a finite-sample analysis of \texttt{DPWideTS}. We look beyond asymptotics solely in $n$, as this will not explain the strong empirical performance of the \texttt{DPWideTS} algorithm compared to others such as \texttt{DPSuffStats}, as shown by~\citep{alabi2022differentially}. In addition, finite-sample analysis will allow us to better understand the \emph{conditions} under which \texttt{DPWideTS} outperforms non-robust algorithms.

The main challenge of analyzing \texttt{DPWideTS} is that the slopes computed from $\matchedindexset$ are correlated.\footnote{We can avoid correlated slopes if we use an `incomplete' version of the algorithm where each data point is only used in one slope, ie. $k=1$. Our general analysis applies to this simplified version, as well as to variants that compute a linear \# of total slopes for computational efficiency.} To deal with correlated slopes,~\cite{Sen68} relies on the properties of \emph{U-statistics}~\citep{Hoe48} for an asymptotic analysis of the non-private Theil-Sen algorithm. 

\begin{mydefinition}[U-statistic for simple linear regression~\citep{Sen68}]
\label{def:u-statistic}
Let $x_1, \ldots, x_n$ satisfy Assumption~\ref{assumption:x-constants}, and let $y_1, \ldots, y_n$ be the corresponding response variables under the model $y_i = \intercept + \slope x_i + e_i$, where $\intercept, \slope \in \reals$ and each $e_i$ is sampled i.i.d from a continuous, symmetric, mean-0 distribution $F_e$. 
The U-statistic takes as input a ``guess'' $\hat{\slope} \in \reals$
for the true slope $\slope$, 
as well as the datapoints $\datatuples$, indexed arbitrarily. Then, the \emph{U-statistic for simple linear regression} is defined as follows.
\begin{align}
    U(\hat{\slope}, \datatuples) 
    = \frac{1}{\numpairs} \sum_{\pairsum} \sign \left( \frac{y_j - y_i}{x_j - x_i} - \hat{\slope} \right),
    \label{eq:u-statistic} 
\end{align}
where $\matchedindexset$ is the set of unordered pairs of datapoints used to compute the slopes (where each pair has distinct $x$ values), 
$\numpairs = |\matchedindexset|$ is the number of such pairs, and $\sign(q) = -1$ if $q < 0$, $0$ if $q=0$, and $1$ if $q > 0$.
For ease of notation, we will 
let $\hat{\slope} = \slope + z$, for some $z \in \reals$, and 
use $U_z$ as shorthand for $U\left(\hat{\slope}, \datatuples\right)$.
\end{mydefinition}

We build on this approach, adapting finite-sample `Berry-Esseen-type' bounds for the convergence of the U-statistic in order to develop a finite-sample convergence bound for the \texttt{DPWideTS} estimate. 

\begin{mytheorem}[Convergence bound for $\betatildewemts$] \label{thm:wemts-bound}
Let $x_1, \ldots, x_n$ satisfy Assumption~\ref{assumption:x-constants} and have empirical variance $\sigma_x^2$. Let $y_1, \ldots, y_n$ be the corresponding response variables under the model $y_i = \intercept + \slope x_i + e_i$, where $\intercept, \slope \in \reals$ and each $e_i$ is sampled i.i.d from a continuous, symmetric, mean-0 distribution $F_e$. 
Let $\betatildewemts = \texttt{DPTheilSen}(\{x_i, y_i\}_{i=1}^n, \eps, (\matchedindexset, \wexpm, \theta, -R, R))$, where $\eps, R, \theta >0$, $\slope \in [-R+\theta, R-\theta]$\removeforsubmission{ as in Assumption~\ref{assumption:range}}, and $\matchedindexset \in \binom{[n]}{2}$ is a set of unordered pairs with distinct x-values. Let $U_z$ be the U-statistic defined with respect to $\matchedindexset$.

Then, there exists a constant $c >0$ such that for sufficiently large $n$, with probability at least $1-p$, $\betatildewemts \in [ \slope - z - \theta, \slope + z + \theta]$ for every $z$ that satisfies the following:
\begin{align*}
    \frac{\mu(z) - \frac{4\ln(4R/p\theta)}{\eps \cdot n}}{\sigma(z)} \geq \Phi^{-1}\left( 1 - \frac{p}{2} + \frac{c}{n^2 \cdot \sigma^3(z)} \right)
\end{align*}
where $\mu(z) = \E[U_z], \sigma^2(z) = \Var[U_z]$, and $\Phi$ is the standard normal cdf.
\end{mytheorem}
\removeforsubmission{
\begin{proof}
See Proof of Thm~\ref{thm:wemts-bound} below in Section~\ref{sec:wemts-proof}.
\end{proof}
}
The quantities $\mu(z)$ and $\sigma^2(z)$ can be further described and evaluated, as shown in Sections~\ref{sec:expectation-variance} and~\ref{sec:special-case}. In the next section, we provide an overview of the analysis and the use of the U-statistic.
However, this result will be easier to interpret when applied to a special setting, as described in Section~\ref{sec:special-case}.


\subsection{Overview of analysis}
\label{sec:overview-of-analysis}
To analyze the performance of \texttt{DPTheilSen}, we further consider the U-statistic from Definition~\ref{def:u-statistic}.
In particular, we can rewrite statistic (\ref{eq:u-statistic}) in the form of a \emph{Kendall's tau statistic}~\citep{Kendall48} that measures the rank correlation between the $x$'s and the residuals from a line of slope $\slope + z$.
\begin{align}
    \label{eq:u-statistic-z-e}
    &U_z 
    = U(\slope + z, \datatuples) 
    = \frac{1}{\numpairsdenom} \sum_{\substack{\{i, j\} \in \matchedindexset \\ i < j}} \sign(x_j - x_i) \cdot \sign \left( e_j - e_i - z \cdot (x_j - x_i) \right) 
\end{align}
Suppose $z=0$.
Then, we define the \emph{null U-statistic} as
\begin{align}
    U_0 
    = U(\slope, \datatuples) 
    = \frac{1}{\numpairsdenom} \sum_{\pairsum} \sign(x_j - x_i) \cdot \sign(e_j - e_i)
\label{eq:u-null-statistic}
\end{align}
 Furthermore, the distribution of $U_0$ matches the null distribution of the Kendall's $\tau$-statistic, which is known to be asymptotically normal~\citep{Kendall48}. Observe that $E_{\es}[U_z] = 0$ since $\es$ are i.i.d. In fact, we have: 
 \begin{fact}[\cite{Sen68}]
 \label{fact:u-expectation-0}
 $\E_{\es}[U_z] = 0$ iff $z = 0$.
 \end{fact}
\removeforsubmission{
We will also point out another fact about the U-statistic, which will be used in our proof of Theorem~\ref{thm:wemts-bound}.
 \begin{fact}[\cite{Sen68}]
 \label{fact:u-nonincreasing}
 For all $z \in \reals$, $U_z$ is a non-increasing function.
 \end{fact}
 }
 Using these properties of the U-statistic, we proceed with the algorithm analysis in two steps.
 \begin{enumerate}[topsep=-1pt,itemsep=0ex,partopsep=0ex,parsep=1ex]
     \item By the utility theorem of the widened exponential mechanism (Theorem~\ref{thm:utility-of-wem}), we show that with high probability, $U\left(\betatildewemts, \datatuples \right) \approx 0$.
 \item Putting together $U\left(\betatildewemts, \datatuples \right) \approx 0$, \removeforsubmission{its non-increasing property,} and the asymptotic normality of $U_z$ for appropriate $z$ (Theorem~\ref{thm:u-statistic-distribution-berry-esseen}), we show that $\betatildewemts \approx \slope$ and characterize the finite-sample convergence of the estimator.
 \end{enumerate}
 
 Note that the second step requires showing that $U_z$, not just $U_0$, is asymptotically normal, and we must characterize the convergence using the quantity $z$ itself. We do so in the next section. 


\subsection{Finite-sample convergence of U-statistic}
\label{sec:finite-sample-convergence}

We develop a finite-sample bound for the absolute value difference between the distribution function of $U_z$ and the standard normal distribution function $\Phi$. To do so, we modify Berry-Esseen bounds for linear functions, such as U-statistics, of i.i.d mean-0 random variables (see, e.g.,~\cite{chen2011normal}). In the case of the U-statistic $U_z$ described in~(\ref{eq:u-statistic-z-e}), the terms $\sign(e_j - e_i - z \cdot (x_j - x_i))$ for $\pairsum$ are non-identically distributed since $x_j - x_i$ may be different for different pairs $(i,j)$. Therefore, we adapt the Berry-Esseen bounds to work for U-statistics of independent, \emph{non-identical} random variables. Our finite-sample convergence bound is stated in Theorem~\ref{thm:u-statistic-distribution-berry-esseen} below. 
\begin{mytheorem}
\label{thm:u-statistic-distribution-berry-esseen}
Let $x_1, \ldots, x_n$ satisfy Assumption~\ref{assumption:x-constants} (ie. not all equal) and let $y_1, \ldots, y_n$ be the corresponding response variables under the model $y_i = \intercept + \slope x_i + e_i$, where $\intercept, \slope \in \reals$ and each $e_i$ is sampled i.i.d from a continuous, symmetric, mean-0 distribution.  
Let $\matchedindexset$ be a set of unordered pairs of datapoints such that each pair has distinct x-values, and let $U_z$ be defined with respect to $\matchedindexset$ as in (\ref{eq:u-statistic-z-e}).
Let $\mu(z) = \E_{\es}[U_z]$, and let $\sigma^2(z) = \Var_{\es}[U_z]$.
Then, for sufficiently large $n$,
\[
    \sup_{t \in \reals} \left \vert \Pr \left[ \frac{ U_z - \mu(z)}{\sigma(z)} \leq t \right] - \Phi(t)
    \right \vert
    = O \left( \frac{1}{n^2 \cdot \sigma^3(z)} \right)
\]
where $\Phi$ is the cdf of a standard normal distribution.
\end{mytheorem}
This result shows that the normalized U-statistic, $U_z$, converges to a standard normal distribution at a rate of 
$O\left(1/(n^2 \cdot \sigma^3(z) )\right)$, where $\sigma^2(z) = \Theta(1/n)$ is the variance of the U-statistic.\footnote{For example, the variance of the U-statistic evaluated for the \splitdesign~is $4/3n + O(1/n)$, as shown in Lemma~\ref{lem:u-endpoints-variance}. The general case is shown in Lemma~\ref{lem:u-bn-statistic-variance}.}

\removeforsubmission{This result is not very easily interpretable in the general case. In the next section, we apply this bound to a special setting and describe how to interpret the terms.}

\section{Evaluating Bound for Special Case} 
\label{sec:special-case}

In this section, we will consider the special setting (called an \splitdesign~\citep{Sen68}) where the x-values are evenly split between the endpoints of an interval of length $\Delta_x$. The \texttt{DPWideTS} algorithm computes $N = \lfloor n/2 \rfloor \cdot \lceil n/2 \rceil$ slopes\footnote{Our analysis extends to efficient versions of this algorithm that compute a linear, rather than quadratic, number of slopes.} from pairs of datapoints on opposite ends of the interval. In addition, we assume that the noise variables are sampled i.i.d from a normal distribution. 
\begin{assumption}
\label{assumption:endpoints}
$x_1, \ldots, x_{\lfloor n/2 \rfloor} = 0$ and $x_{\lfloor n/2 \rfloor + 1}, \ldots, x_n = \Delta_x$.
\end{assumption}
\vspace{-1em}
\begin{assumption}
\label{assumption:normal-errors}
    $F_e = \calN(0, \sigma_e^2)$.
\end{assumption} 

\removeforsubmission{
In Theorem~\ref{thm:wemts-bound-endpoints} below, we state a convergence bound for this special case. The bound relies on the general bound of Theorem~\ref{thm:wemts-bound} along with evaluations of the expectation and variance of the U-statistic.
}
\removeforsubmission{We are able to simplify the expressions for the expectation and variance because in this special case, all of the slopes are drawn from an identical distribution, yet we still require Theorem~\ref{thm:wemts-bound} because the slopes are correlated.}

\removeforsubmission{
\begin{thmn}[\ref{thm:wemts-bound-endpoints}]
Let $x_1, \ldots, x_n$ be at two endpoints of an interval of size $\Delta_x$ such that they satisfy Assumption~\ref{assumption:endpoints}, and let $e_1, \ldots, e_n$ be drawn i.i.d from $F_e = \calN(0, \sigma_e^2)$ according to Assumption~\ref{assumption:normal-errors}. Let $y_1, \ldots, y_n$ be the corresponding response variables under the model $y_i = \intercept + \slope x_i + e_i$, where $\intercept, \slope \in \reals$.
Let $\betatildewemts = \texttt{DPTheilSen}(\{x_i, y_i\}_{i=1}^n, \eps, (\wexpm, \theta, -R, R))$, where $\eps, R, \theta >0$, and $\slope \in [-R+\theta, R-\theta]$ as in Assumption~\ref{assumption:range}.
Let $\tau$ be defined as follows.
\begin{align*}
    \tau = \Phi^{-1} \left( 1 - \frac{p}{8} \right) \cdot \sqrt{\frac{4}{3n}} + \frac{8\ln (4R/p\theta)}{\eps n} 
\end{align*}
where $\Phi^{-1}$ is the inverse standard normal distribution function.
Then, we have that for sufficiently large $n$, where for some constant $c$, $n \geq c \left( \log(R/p\theta) / \eps \sqrt{\log(1/p)} \right)^2$, and sufficiently small $\tau$,  we have that with probability at least $1-p$, 
\begin{align*}
\betatildewemts \in [\slope - z - \theta, \slope + z + \theta] \\
\text{for}~ z = \frac{2\sqrt{\pi} \cdot \sigma_e}{\Delta_x} \cdot \left( \tau + O\left(\tau^{3/2}\right) \right).
\end{align*}
\end{thmn}
}
\removeforsubmission{
\begin{proof}
See Proof of Thm~\ref{thm:wemts-bound-endpoints} below.
\end{proof}
}

Table~\ref{tab:comparison-endpoints} shows the \texttt{DPWideTS} convergence bound for this special setting, along with 
\removeforsubmission{In order to provide intuition for this result, w}
convergence bounds for the non-private algorithms OLS and Theil-Sen and the private algorithms \texttt{DPSuffStats} and \texttt{DPOLSExp}~\citep{alabi2022algorithmic} (these were described briefly in Section~\ref{sec:linreg}), all under the same assumptions.

\begin{table*}[hbtp!]
\centering
  \begin{tabular}{ | l | c |  c | } 
    \hline 
    & & \\ [-.2ex]
    Estimator & Size of $1-p$ Convergence Bound & Constraints \\ [1ex]
    \hline 
    & & \\ [-.2ex]
    OLS &
    $ 2 \cdot \frac{\sigma_e}{\Delta_x} \cdot \frac{\nonprivconv{p/2}}{\sqrt{n}}$ & \\ [1.5ex]
    \hline
    & & \\ [-.2ex]
    Theil-Sen \small \citep{Sen68} & $ 2 \cdot \sqrt{\frac{\pi}{3}} \cdot \frac{\sigma_e}{\Delta_x} \cdot \frac{\nonprivconv{p/4}}{\sqrt{n}} $
    & \\ [1.5ex]
    \hline
    & & \\ [-.2ex]
    \texttt{DPSuffStats}  &
    $ 2 \cdot \frac{\sigma_e}{\Delta_x} \cdot \frac{\nonprivconv{p/6}}{\sqrt{n}} \cdot \left( 1 + \tau \right) + \tau \left( 1 + \tau + |\slope| \right)$,
    & 
    \small $p \in (3\exp (-\eps n \Delta_x^2 / 12 ), 1 )$ \\
    \small \citep{alabi2022algorithmic} & $\tau \approx \frac{12(1-1/n)r^2 \log(3/p)}{\eps \cdot n \cdot \Delta_x^2}$ &
    \\ [1.5ex]
    \hline
    & & \\ [-.2ex]
    \texttt{DPOLSExp} \small \citep{alabi2022algorithmic} &
    $ 2 \cdot \frac{\sigma_e}{\Delta_x} \cdot \frac{\nonprivconv{p/4}}{\sqrt{n}} + \sqrt{\frac{64 R \cdot (3\log(3) + \log(2/p))}{\eps n^2 \Delta_x^2}} $ &
    \\ [1.5ex]
    \hline
    & & \\ [-.2ex]
        \texttt{DPWideTS}  & 
    $2 \cdot \sqrt{\frac{\pi}{3}} \cdot \frac{\sigma_e}{\Delta_x} \cdot \left( \frac{\nonprivconv{p/8}}{\sqrt{n}} + \frac{4 \sqrt{3} \ln (4R/p\theta)}{\eps \cdot n} \right) + \theta$
    & \small $p \in (0, 4R/\theta)$, \\
     \small (Thm~\ref{thm:wemts-bound-endpoints}) & & \small $\frac{\sigma_e}{\Delta_x}$ suff. small, $n$ suff. large\\  [1.5ex]
    \hline
  \end{tabular} 
  \\ [1.5ex]
\caption{High-probability ($1-p$) convergence bounds for estimators for special setting (Assumptions~\ref{assumption:endpoints} and~\ref{assumption:normal-errors}). Note that $r$ is a hyperparameter range for both the input $x_i, y_i$ datapoints, while $R$ is a range for the output estimate of the slope.}
\label{tab:comparison-endpoints}
\end{table*}

The first two bounds in the table correspond to the non-private algorithms OLS and Theil-Sen. Both of these non-private algorithms
have a convergence rate of $O(\sigma_e \cdot \nonprivconv{p} / \Delta_x \cdot \sqrt{n} )$. Looking at the leading constant, we see that Theil-Sen nearly recovers the accuracy of OLS, up to a factor of $\sqrt{\pi/3}$. The Theil-Sen bound has the term $\nonprivconv{p/4}$ instead of $\nonprivconv{p/2}$, which comes from the convergence of the U-statistic to a standard normal distribution at a rate of $O(1/\sqrt{n})$ (Theorem~\ref{thm:u-statistic-distribution-berry-esseen}).

The bounds for
\texttt{DPSuffStats} and \texttt{DPTheilSen} have the same constant factors for the highest order term as OLS and Theil-Sen, respectively, but they also contain two main differences from their non-private counterparts. First, the DP estimators have constant factor changes in $p$ in the $\nonprivconv{p}$ terms compared to those in the non-private bounds; this comes from the DP estimators taking a union bound over both the sampling and privacy error, which splits up the failure probability $p$ further. Note that $\nonprivconv{p} = \Theta(\sqrt{\log(/p)})$ for small $p$, so these constant factor changes in $p$ are not impactful.

Second, the DP estimators include lower order terms corresponding to the noise due to privacy,\footnote{For example, the $\ln(R/p\theta)/\eps n$ term for \texttt{DPWideTS} corresponds to the noise due to privacy, which is $O(1/\eps n)$ and thus overshadowed by the sampling error $O(1/\sqrt{n})$ for sufficiently large $n$.}
which
provide insight into the relative performance of the DP algorithms in practical regimes.
For example, we can see that \texttt{DPSuffStats}'s lower order term $\tau$ is $O(r^2/\eps \cdot n \cdot \Delta_x^2)$; the $\eps n \Delta_x^2$ quantity in the denominator was highlighted in the empirical work of~\cite{alabi2022differentially} as important for the performance of the algorithm but lacked theoretical backing prior to our work.\footnote{Alabi et al. specificially looked at the quantity $\eps \cdot n \cdot \sigma_x^2$,  where $\sigma_x^2$ is the empirical variance of the constants $x_1, \ldots, x_n$, but they did not provide theoretical basis for the importance of this quantity. In our special setting, $\Delta_x^2 = 4 \sigma_x^2$.} Meanwhile,
the lower order term in \texttt{DPOLSExp} has the quantity $\sqrt{\eps} n \Delta_x$ in the denominator, which indicates that it performs better than \texttt{DPSuffStats} for small $n$ and for small $\eps$. The lower order term in \texttt{DPWideTS} has the quantity $\eps n \Delta_x$ in the denominator, which indicates that it performs better than \texttt{DPSuffStats} for small $\Delta_x$, but not as well as \texttt{DPOLSExp} in the small $\eps$ regime.

Finally, we can compare the \texttt{DPSuffStats}, \texttt{DPOLSExp}, and \texttt{DPWideTS} bounds in terms of their dependence on hyperparameters. 
While \texttt{DPSuffStats} has a quadratic dependence on $r$, the range of the input datapoints, the other two algorithms avoid any dependence on the range of the inputs. Instead, \texttt{DPOLSExp} has a square root dependence on the range $R$ of the output estimate, while \texttt{DPWideTS} has a milder logarithmic dependence on $R$. The bound  for \texttt{DPWideTS} also includes the parameter $\theta$ (corresponding to the granularity of outputs in \texttt{DPWide}). In Section~\ref{sec:choosing-theta}, we discuss how to set this parameter, addressing an open question of~\citep{alabi2022algorithmic}.

\removeforsubmission{
\subsection{Choosing the widening parameter, $\theta$}
\label{sec:choosing-theta}
Theorem~\ref{thm:wemts-bound-endpoints} offers insight on how to set $\theta$ in the \texttt{DPWideTS} algorithm, which was an open question raised by~\cite{alabi2022differentially}. For fixed $p, \eps, R, \sigma_e, \Delta_x$, and $n$ that satisfy the conditions stated in the Theorem,
we can set $\theta$ to minimize the bound as follows:
\begin{align*}
 \theta \approx \max \left( \frac{\sigma_e \cdot \ln (R \eps n \Delta_x / p \sigma_e)}{\eps n \Delta_x}, \frac{R}{p} \exp(-\eps n) \right)
\end{align*}
The first term in the max comes from allowing the two terms involving $\theta$ in Thm.~\ref{thm:wemts-bound-endpoints} to be approximately equal, whereas the second term in the max comes from upper bounding $O(\ln(R/p\theta)/\eps n)$ by some constant. The reason for having both terms is to account for both widespread and concentrated slopes.\footnote{Handling the case of concentrated slopes was~\cite{alabi2022differentially}'s original motivation for designing the widened exponential mechanism.}
The factor $\sigma_e / (n \Delta_x)$ in the first term corresponds to the standard deviation of the slopes computed by \texttt{DPWideTS}; when the slopes are highly concentrated, the first term becomes small.
The second term, however, is independent of $\sigma_e$ and $\Delta_x$, which allows $\theta$ to remain bounded away from $0$ and prevents a blowup in the convergence bound.
Note that $R, \eps, n$ and $p$ are known in practice; if the experimental design suggests that the slopes may be concentrated (e.g. if the x-values are located at one of two endpoints of an interval as in this special case), our analysis suggests setting $\theta$ to scale with the second term.
}


\section{Confidence Intervals for DPWideTS}
In addition to the DP point estimators, we also consider DP confidence intervals for the linear regression slope $\slope$. 
\removeforsubmission{
\JS{check this}
\begin{definition}[Non-parametric confidence intervals for simple linear regression.]
Let $\bd = \datatuples$ be a dataset real-valued pairs, where $x_1, \ldots, x_n$ are fixed values that are not all equal, and $y_1, \ldots, y_n$ are the corresponding response variables under the model $y_i = \intercept + \slope x_i + e_i$, where $\intercept, \slope \in \reals$ and each $e_i$ is sampled i.i.d from a continuous, symmetric, mean-0 distribution $F_e$.
Let $I_{\reals}$ be the set of all intervals in $\reals$, and let $M_{\textrm{nonpriv}}: \dataspacereal \times \calH \rightarrow I_{\reals}$ be a deterministic mechanism that outputs an interval. For any $\alpha_1 \in (0,1)$, $M_{\textrm{nonpriv}}$ outputs a \emph{$(1-\alpha_1)$-confidence interval} for the slope $\slope$ if for all hyperparams $\in \calH$, and sufficiently large $n$, 
\begin{align*}
    \Pr_{\es \sim F_e} \left[ \slope \in M_{\textrm{nonpriv}}(\bd, \text{hyperparams}) \right] \geq 1-\alpha_1
\end{align*}
where the probability is over the randomness of the dataset $\bd$. 
Next, let $M_{\textrm{DP}}: \dataspacereal \times \posreals \times \calH \rightarrow I_{\reals}$ be a randomized $\eps$-DP mechanism that outputs an interval. For any $\alpha \in (0,1)$, $M_{\textrm{DP}}$ outputs a $(1-\alpha)$-\emph{DP confidence interval} for the slope $\slope$ if for all privacy loss parameters $\eps \in \posreals$, hyperparams $\in \calH$, and sufficiently large $n$,
\begin{align*}
    \Pr_{\substack{\es \sim F_e \\ M_{\textrm{DP}}}} \left[ \slope \in M_{\textrm{DP}}(\bd, \eps, \text{hyperparams}) \right] \geq 1-\alpha
\end{align*}
\end{definition}
}
\removeforsubmission{Now, we briefly describe the \emph{non-private} confidence interval for \texttt{Theil-Sen}. This} 
We start with the non-private Theil-Sen estimator, described in~\cite{Sen68}. This estimator returns the $(1/2-b)$th and $(1/2+b)$th quantiles of the set of \texttt{Theil-Sen} slopes, $\bs$, where $b$ is computed according to the distribution of the corresponding U-statistic.\justforsubmission{\footnote{See the Appendix for more details on this interval estimator and a proof of its validity.}} 
\removeforsubmission{(or, as described below, its approximation by a normal distribution).}

\removeforsubmission{
\begin{definition}
Let $\matchedindexset$ be a set of $\numpairs$ pairs of points, and let $U_0$ be the corresponding null U-statistic defined in Definition~\ref{def:u-statistic} Equation (\ref{eq:u-null-statistic}). For the corresponding set of slopes $\bs$, let $\fs$ be the empirical distribution function for $\bs$. Let $\sigma^2(0) = \Var_{\es}[U_0]$ as evaluated in Corollary~\ref{lem:u-null-statistic-variance}.
For any $\alpha_1 \in (0,1)$, let  $b$ is defined as: 
\begin{align*}
    b:= \frac{1}{2} \cdot \Phi^{-1} \left(1-\frac{\alpha_1}{8} \right) \cdot \sigma(0)
\end{align*}
and where $\Phi$ is the standard normal cdf. 
Then, the \emph{non-private $1-\alpha_1$-Theil-Sen confidence interval} is $[\hat{\slope}_L, \hat{\slope}_U]$, where
$\hat{\slope}_L = \fsinv(1/2 - b)$ and $\hat{\slope}_U = \fsinv(1/2 + b)$.
\end{definition}
The proof of this interval's validity can be found in Lemma~\ref{lem:ts-ci-nonpriv-validity}.
}

\removeforsubmission{\subsection{Union-bound confidence interval for \texttt{DPWideTS}}}
Now, we consider the differentially private confidence interval for $\slope$, \texttt{DPTSCI}, which we describe in Algorithm~\ref{alg:dpts-ci} \justforsubmission{(See Appendix)}. This algorithm is similar to the \texttt{DPTheilSen} point estimator (Alg.~\ref{alg:ts}), except that we replace the call to \texttt{DPMed} with a call to an $\eps$-DP confidence interval algorithm for the median, \texttt{DPMedCI}. 

\removeforsubmission{
\begin{algorithm}[hbtp!]
  \KwData{$\bd = \datatuples \in \dataspacereal$}
  \KwPrivacyparams{$\eps \in \nonnegreals$}
  \KwHyperparams{$\matchedindexset \in {[n] \choose 2}$, \texttt{DPMedCI}, \text{hyperparams} $\in \calH$}

  $\slopes = \{\}$

  \For{each $\{i,j\} \in \matchedindexset$ (such that $x_i \neq x_j$)} {
    
        $s = (y_j - y_i)/(x_j - x_i)$
    
    Add $s$ to $\slopes$

  }
  Let $k = \max_{i \in [n]} \{\#j \in [n]: \{i, j\} \in \matchedindexset\}$
  
  $[\betatildets_L, \betatildets_U] = \texttt{DPMedCI}\left(\slopes, \eps/k, \text{hyperparams}\right)$ 
  
  \Return $[\betatildets_L, \betatildets_U]$
  \caption{\texttt{DPTSCI}: $\eps$-DP Algorithm}
  \label{alg:dpts-ci}
\end{algorithm}
\begin{lemma}
\label{lem:dptsci-privacy}
Algorithm~\ref{alg:dpts-ci} is $\eps$-DP.
\end{lemma}
}

In Alg.~\ref{alg:dpwem-ci}, we describe one possible \texttt{DPMedCI} algorithm: \texttt{DPWideCI(Union)}. This algorithm is based on a nonparametric confidence interval for the median designed by~\cite{DGMSS21}. The idea is to run the \texttt{DPWide} point estimator twice such that with high probability, the two estimates capture the true slope $\slope$. The \texttt{DPWideCIUnion} algorithm does so by outputting an interval that contains the non-private interval with high probability, while the \texttt{DPWideCI} algorithm outputs a tighter interval via more nuanced coverage analysis. While~\cite{DGMSS21} assume that the inputs to the confidence interval mechanism are sampled i.i.d. from a population distribution, here we only assume that we have a uniform bound on the convergence of the empirical distribution $\fs$ to the standard normal distribution (which we develop in Thm.~\ref{thm:u-statistic-distribution-berry-esseen}).


\begin{algorithm}[h!]
  \KwData{$\bs = (s_1, \ldots, s_\numpairs) \in \reals^\numpairs$}
  \KwPrivacyparams{$\epsinput$}
  \KwHyperparams{$\alphainput, r_{\alpha} \in (0,1), \graninput, \rangeinput, \texttt{Union} \in \{0, 1\}$}
  
  \If{\texttt{Union}}{

  Let $\alpha_1 = r_\alpha \cdot \alpha$ and $\alpha_2 = (1-r_\alpha) \cdot \alpha$
  
  Let $b = \frac{1}{2} \cdot \Phi^{-1} \left(1-\alpha_1/8 \right) \cdot \sigma(0)$ ~\tcp{\small $\sigma^2(0)$ is the variance of the null U-statistic, $U_0$, corresponding to the set of slopes $\bs$. As shown in Corollary~\ref{lem:u-null-statistic-variance}, it does not depend on the data.}
  
  Let $c = 2\ln(4R/\alpha_2 \theta)/(\eps \numpairs)$
  
  Set $q_L = 1/2 - b - c$ and $q_U = 1/2 + b + c$
  }
  \Else{
 $q_L, q_U = \texttt{ComputeExpMechCITargets}(\bs, \eps, \alpha, -R, R, \theta)$ ~~~\tcp{\small Algorithm~\ref{alg:exp-mech-compute-targets}}  
 }
  
  $\privCIL{\alpha}(\bs) = \wexpm(\bs, \eps/2, ( q_L, \range, \theta))$ ~~~\tcp{\small Algorithm~\ref{alg:wem}}
  
  $\privCIU{\alpha}(\bs) = \wexpm(\bs, \eps/2, (q_U, \range, \theta))$
    
\Return $[\privCIL{\alpha}(\bs) - \theta, \privCIU{\alpha}(\bs) + \theta]$
  \caption{$\texttt{DPWideCI(Union)}$:  
  $\eps$-DP Algorithm}  
  \label{alg:dpwem-ci}
\end{algorithm}

Let \texttt{DPWideTSCIUnion} = \texttt{DPTSCI} ($\bd$, $\eps$, $\matchedindexset$, \texttt{DPWideCIUnion}, ($\alpha$, $r_\alpha$, $\theta$, $-R$, $R$, \texttt{Union}=1)), where \texttt{DPWideCIUnion} is defined in Algorithm~\ref{alg:dpwem-ci}. 
Our main theorem for \texttt{DPWideTSCIUnion}, stated \removeforsubmission{below}\justforsubmission{in the Appendix}, describes the validity and width of this confidence interval.
The proof of the validity relies on the validity of the non-private interval (Lemma~\ref{lem:ts-ci-nonpriv-validity}), along with the utility of \texttt{DPWide} (Thm.~\ref{thm:utility-of-wem}). \removeforsubmission{The proof of the width is similar to the proof of Thm.~\ref{thm:wemts-bound}.}
\removeforsubmission{
\begin{mytheorem}[Validity and width of \texttt{DPWideTSCIUnion}] 
\label{thm:wemts-ci-bound}
Let $x_1, \ldots, x_n$ satisfy Assumption~\ref{assumption:x-constants}, and let $y_1, \ldots, y_n$ be the corresponding response variables under the model $y_i = \intercept + \slope x_i + e_i$, where $\intercept, \slope \in \reals$ and each $e_i$ is sampled i.i.d from a continuous, symmetric, mean-0 distribution $F_e$. For $\alpha \in (0,1)$, $r_\alpha \in (0,1)$, $\eps, R, \theta >0$, and $\slope \in [-R+\theta, R-\theta]$ as in Assumption~\ref{assumption:range},
let $[\betatildewemts_L, \betatildewemts_U] =$ 
$\texttt{DPTSCI}(\{x_i, y_i\}_{i=1}^n, \eps, \texttt{DPWideCIUnion}, (\alpha, r_\alpha, \theta, -R, R, \texttt{Union}=1))$. Then,
the interval $[\betatildewemts_L, \betatildewemts_U]$ is a $1-\alpha$-confidence interval for the true slope $\slope$ over the randomness in $\datatuples$ and \texttt{DPWideTSCIUnion}.

In addition, 
for a set $\matchedindexset$ of $\numpairs$ unordered pairs of datapoints, let $U_z$ be defined as in Definition~\ref{def:u-statistic}, Equation (\ref{eq:u-statistic-z-e}). 
Then, there exists a constant $c$ such that for sufficiently large $n$ \JS{add condition}, we have that with probability at least $1-\alpha$, $ \betatildewemts_U - \betatildewemts_L < 2z + 2\theta$ for all $z$ that satisfies the following.
\begin{align*}
    \frac{-\mu(z) - \frac{8\ln(2R/\alpha \theta)}{\eps \cdot n}}{\sigma(z)} \geq 2 \cdot \Phi^{-1}\left( 1 - \frac{\alpha}{2} + \frac{c}{n^2 \cdot \sigma^3(z)} \right)
\end{align*}
where $\mu(z) = \E[U_z], \sigma^2(z) = \Var[U(z)]$, and $\Phi$ is the standard normal cdf.
\end{mytheorem}
}
The width of the \texttt{DPWideTSCIUnion} confidence interval is approximately twice the size of the convergence bound of the point estimator (Thm.~\ref{thm:wemts-bound}), which we expect even in the non-private setting. However, there is an additional factor of $4$ in the term corresponding to privacy noise ($\ln(2R/\alpha \theta)$. This can be improved using a more nuanced coverage analysis\removeforsubmission{, as we will explore in the next section}.

\removeforsubmission{\subsection{Tighter Confidence Interval}}
To improve the width of the confidence interval, we apply the more sophisticated approach from~\cite{DGMSS21}. In particular, we call $\texttt{DPWideCI}$ with the \texttt{Union} flag set to 0, so that $q_L, q_U$ are computed using \texttt{ComputeExpMechCITargets} (Alg.~\ref{alg:exp-mech-compute-targets}). We offer a brief overview here:
For a given $t_L, t_U \in [\numpairs]$, $\eps \in \posreals$, let $\tilde{L_\alpha}(\bs, t_L) = \wexpm(\bs, \eps/2, (t_L/\numpairs, \theta, -R, R))$ and $\tilde{U_\alpha}(\bs, t_U) = \wexpm(\bs, \eps/2k, (t_U/\numpairs, \theta, -R, R))$. 
As before, the goal is to control the probability that the interval $[\tilde{L_\alpha}(\bs, t_L) - \theta, \tilde{U_\alpha}(\bs, t_U) + \theta]$ fails to contain the true slope $\slope$. In particular, for $\alpha \in (0, 1)$, we would like to find the target ranks $t_L$ and $t_U$ closest to $\numpairs/2$ such that
\begin{align}
    \Pr_{\bs, \wexpmci} \left[ \tilde{L_\alpha}(\bs, t_L) - \theta > \slope \right] \leq \alpha/2,  ~~\text{and} 
    \Pr_{\bs, \wexpmci} \left[ \tilde{U_\alpha}(\bs, t_U) + \theta < \slope \right] \leq \alpha/2 \label{eq:em-better-analysis}
\end{align}
In Algorithm~\ref{alg:exp-mech-compute-targets}, we find these target ranks by first computing the probabilities above for all possible $t_L$ and $t_U$'s, and then by numerically searching for the target ranks closest to $\numpairs/2$ such that the probabilities above are both within $\alpha/2$. This search can be implemented more efficiently by noting that $t_L$ is greater than or equal to $\numpairs \cdot (1/2 - b - c)$ as defined in Algorithm~\ref{alg:dpwem-ci}, and similarly $t_U$ is less than or equal to $\numpairs \cdot (1/2 + b + c)$.
\justforsubmission{ This also tell us that the width of the interval will always be the same or smaller than that of the \texttt{DPWideTSCIUnion} interval (Thm.~\ref{thm:wemts-ci-bound}).}

\removeforsubmission{
\begin{algorithm}[hbtp!]
  \KwData{$\bs = (s_1, \ldots, s_{\numpairs})$}
  \KwInput{$\epsinput, \alphainput, \rangeinput, \graninput$}
  
  \For{$t_L \in \mathbb{N}, 1 \leq t_L \leq \numpairs/2 $}{
  
  $\gamma_{t_L}=1 - F_U \left( 1 - 2(t_L -1)/\numpairs \right) + \sum_{m = t_L}^{n} F'_U \left( 1- 2m/\numpairs \right) \cdot \frac{R}{\theta}\exp(- (m-t_L) \cdot \eps / 2 )$
  } \tcp{$F_U, F'_U$ are the CDF and PDF of the null U-statistic corresponding to $\bs$}
  $q_L = \max_{t_L \in \mathbb{N}, 1 \leq t_L < \lceil \numpairs/2 \rceil} \{t_L : \gamma_{t_L} \leq \alpha/2 \} / \numpairs$
  
  \For{$t_U \in \mathbb{N}, \numpairs/2 \leq t_U < \numpairs$}{
  
  $\gamma_{t_U} = F_U\left(1-2 t_U / \numpairs \right) + \sum_{m = 1}^{t_U} F'_U\left(1-2m/\numpairs\right) \cdot \frac{R}{\theta} \exp(- (t_U-m) \cdot \eps / 2 ) $
  }
  $q_U = \min_{t_U \in \mathbb{N}, \lceil \numpairs/2 \rceil \leq t_U < \numpairs} \{j: \gamma_{t_U} \leq \alpha/2 \} / \numpairs$
  
  \Return $q_L, q_U$
  
  \caption{\texttt{ComputeExpMechCITargets}}  \label{alg:exp-mech-compute-targets}
\end{algorithm}
}
\removeforsubmission{
As the outputs $q_L, q_U$ of \texttt{ComputeExpMechCITargets} are $ \geq 1/2 - b - c$ and $\leq 1/2 + b + c$, where $b, c$ are defined in \texttt{DPWideCIUnion} algorithm, the width of the interval will always be less than or equal to that of the \texttt{DPWideTSCIUnion} interval (Thm.~\ref{thm:wemts-ci-bound}).
}
The validity analysis for Alg.~\ref{alg:exp-mech-compute-targets} (Thm.~\ref{thm:wemtsci-validity-better}) relies on characterizing the distribution of the rank of the population median in the dataset. \removeforsubmission{Drechsler et al. used the fact that for i.i.d. data, the rank of the population median in the dataset follows a binomial distribution. } In our setting, where the slopes $\bs$ computed by \texttt{DPTSCI} are not necessarily i.i.d, we\removeforsubmission{instead characterize the rank of the population median (ie. the true slope $\slope$)} \justforsubmission{do so} via the distribution of the corresponding U-statistic (Thm.~\ref{thm:u-statistic-distribution-berry-esseen}).
\removeforsubmission{
\begin{mytheorem}
\label{thm:wemtsci-validity-better}
Let $x_1, \ldots, x_n$ satisfy Assumption~\ref{assumption:x-constants}, and let $y_1, \ldots, y_n$ be the corresponding response variables under the model $y_i = \intercept + \slope x_i + e_i$, where $\intercept, \slope \in \reals$ and each $e_i$ is sampled i.i.d from a continuous, symmetric, mean-0 distribution $F_e$. For $\alpha \in (0,1)$, $r_\alpha = 1/2$ (as a default, since it won't be used), $\eps, R, \theta >0$, and $\slope \in [-R+\theta, R-\theta]$ as in Assumption~\ref{assumption:range},
let $[\betatildewemts_L, \betatildewemts_U] =$ 
$\texttt{DPTSCI}(\{x_i, y_i\}_{i=1}^n, \eps, \texttt{DPWideCI}, (\alpha, r_\alpha, \theta, -R, R, \texttt{Union}=0))$. Then,
the interval $[\betatildewemts_L, \betatildewemts_U]$ is a $1-\alpha$-confidence interval for the true slope $\slope$ over the randomness in $\datatuples$ and \texttt{DPWideTSCI}.
\end{mytheorem}
}

\section{Conclusion}
In this work, we analyze the theoretical privacy and utility guarantees of \texttt{DPTheilSen}. We provide finite-sample convergence bounds, offer insight into hyperparameter selection, and show how to produce differentially private confidence intervals. Future work should analyze the optimality of these algorithms and extend the confidence intervals to multivariate settings.

\acks{We thank Daniel Alabi, Audra McMillan, and Adam Smith for discussions that motivated this work. We also thank Cynthia Dwork and
James Honaker for helpful conversations at an early stage of this
work, and Ryan Cumings-Menon and others at the Fields Institute Privacy Workshop in Summer 2022
for helpful comparisons with related work on robust regression. 

SV was supported by a Simons Investigator
Award and JS was supported by the Columbia Data Science Institute Fellowship. Both authors were supported by Cooperative Agreement CB20ADR0160001 with the Census
Bureau. The views expressed in this paper are those of the authors and not those of
the U.S. Census Bureau or any other sponsor.

}

\bibliography{bib}

\newpage

\appendix





\section{Related Work, continued}
\label{sec:related-work-cont}

We begin by describing related work in more detail. Several recent works have leveraged the connection between robustness and differential privacy to design privacy-preserving algorithms for specific statistical tasks, affirming~\cite{DworkL09}'s initial findings that robust estimators are a valuable starting point for accurate DP algorithms. For example,~\cite{CKSBG19} find that robust estimators perform better than parametric estimators under differential privacy, even when the data come from a parametric model, but they focus on hypothesis testing and do not provide a theoretical utility analysis.~\cite{avella2021privacy} designs DP estimators based on bounded influence M-estimators, including DP tests for Wald, score, and likelihood ratio tests; they provide asymptotic analysis of the statistical guarantees, while our work focuses on finite-sample analysis. 

More generally, recent work has significantly advanced our understanding of the connection between robust and private algorithms.
Work by~\cite{asi2023robustness} establishes a tight connection between privacy and robustness, providing a 
black-box transformation from optimal robust to optimal differentially private algorithms. They also design and analyze estimators for DP linear regression for high-dimensional tasks under assumptions of Gaussianity. 
Work by~\cite{hopkins2023robustness} shows how to implement this black-box transformation in a computationally efficient manner via the sum-of-squares method. While these methods provide important general toolkits for DP algorithm design, they do not provide finite-sample bounds for specific estimators such as \texttt{DPTheilSen}.

Many recent works in the DP literature have considered linear regression, but most focus on empirical or asymptotic analysis rather than finite-sample guarantees.~\cite{Sheffet17} considered differentially private ordinary least squares (OLS) methods and corresponding DP confidence intervals, but unlike our work, these methods assume normality of errors, require input data bounds, and satisfy approximate, rather than pure, DP.~\cite{Wang18} studied private ridge regression and considered DP confidence intervals, but these methods require consuming additional privacy budget for estimating Hessians.~\cite{Barrientos:2017} and~\cite{Evans:2021} use the subsample-and-aggregate framework, but their approaches rely on normality assumptions or normal approximations that only hold for large $n$.~\cite{Bernstein:2018} consider a Bayesian approach, but unlike our work, they require a prior on the distribution of both the regression coefficients and the independent variables.  

Finally, recent studies have further highlighted the need for more usable DP algorithms for basic statistical tasks such as simple linear regression. A study of data practitioners' use of DP by~\cite{sarathy2023don}
calls for algorithms with minimal hyperparameters and assumptions, as well as useful uncertainty measures, that will enable data analysts to navigate the constraints of the DP analysis process. Furthermore, a study by~\cite{barrientos2021feasibility} evaluates a range of DP linear regression algorithms in terms of feasibility for real-world use. They develop criteria around evaluating algorithms, including: assumptions that align with practical applications, ease of implementation, computational efficiency, minimal tuning parameters, and accompanied by uncertainty estimates. This study finds that suprisingly few algorithms in the DP literature satisfy these criteria. Our work on developing the design and analysis of \texttt{DPTheilSen} makes this family of algorithms measure up across all of these categories, suggesting that it will be useful for practical deployments of DP.

\section{Widened Exponential Mechanism (\texttt{DPWide})}

In this section, we describe the \emph{widened exponential mechanism} for quantile estimation designed by~\cite{alabi2022differentially}, which is a variant of the standard exponential mechanism~\citep{McSherryT07} described below.

\begin{mydefinition}[Exponential Mechanism~\citep{McSherryT07}] 
The exponential mechanism is defined with respect to a utility function $u$, which maps (data set, output) pairs to real values. Given dataset $\bs$ and range of possible outputs, $[-R, R]$, the exponential mechanism outputs $r \in [-R, R]$ with probability density proportional to $\exp\left(\frac{\eps u(\bs, r)}{2 \Delta u}\right)$, where $\Delta(u)=\sup_{r\in[-R,R]}\sup_{\bs \sim \bs'} |u(\bs, r) - u(\bs', r)|$. The exponential mechanism is $\eps$-DP.
\end{mydefinition}

The standard instantiation of the exponential mechanism for quantile estimation (used, e.g., in~\cite{Smi11}) uses a utility function that assigns a score to output $r$ based on how far $r$ is \emph{in rank} from the desired quantile of $\bs$. For all outputs $r$ within the range $[-R, R]$, and for a target quantile $q \in (0,1)$, the standard utility function is:
\[
u(\bs, r) = - \lfloor \left \vert
\#\text{below } r - n \cdot q 
\right\vert \rfloor 
\]
where 
\#below $r$ denotes the number of datapoints in $\bs$ that are 
less than or equal to $r$ in value. Note that this utility function assigns the same utility score to every output $r$ in the interval between two data points. 

One issue with this mechanism is that when the output space is the real line and the data is highly concentrated, the mechanism may not place enough probability density near the target quantile. To mitigate this issue,~\cite{alabi2022differentially} design a variation on the standard utility function. For widening parameter $\theta>0$, and target quantile $q \in (0,1)$, the widened utility function is:
\[u(\bs, r) = - \lfloor\min_{w \in [r-\theta, r+\theta]}
\left \vert
\#\text{below } w - n \cdot q
\right\vert \rfloor \]
This utility function provides a lower bound on the probability density the mechanism assigns around the target quantile.
Following~\cite{alabi2022differentially}, an efficient implementation of the $\theta$-widened exponential mechanism for the median is given in Algorithm~\ref{alg:wem}.

\begin{algorithm}[hptb!]
  \KwData{$\slopes = (s_1, \ldots, s_\numpairs) \in \reals^\numpairs$}
  
  \KwPrivacyparams{$\eps \in \nonnegreals$}
  
  \KwHyperparams{$q \in (0,1), -R, R \in \reals^2; \graninput$}
  
  Sort $\slopes$
  
  \Comment{Clip $\slopes$ to the range $[-R, R]$ and insert space $\theta$ around the $q$th quantile:}
    

    
  
  \For{$i \in [1, \lfloor \numpairs q \rfloor ]$}{
    $\slopes[i] = \min(\max(-R, \slopes[i] - \theta), R)$
   }
   
   \For{$i \in [\lfloor \numpairs q \rfloor + 1, \numpairs]$}{
    
    $\slopes[i] = \max(\min(R, \slopes[i] + \theta), -R)$
  }
  
  Insert $-R$ and $R$ into $\slopes$ and set $\numpairs = \numpairs+2$
  
  Set $\textrm{maxNoisyScore} = -\infty$
  
  Set $\textrm{argMaxNoisyScore} = -1$

  \For{$i \in [2, \numpairs)$} {
    
    
    $\textrm{score} = \log(\slopes[i] - \slopes[i-1]) -\frac{\eps}{2} \cdot \lfloor | i - \numpairs q |\rfloor $
    
    $Z \sim \textrm{Gumbel}(0,1)$ 
    
    $\textrm{noisyScore} = \textrm{score} + Z$
    
    \If {$\textrm{noisyScore} > \textrm{maxNoisyScore}$} {
        $\textrm{maxNoisyScore}= \textrm{noisyScore}$
        
        $\text{argMaxNoisyScore} = i$
        }
  }
  
  $\text{left} = \slopes[\text{argMaxNoisyScore}-1]$
  
  $\text{right} = \slopes[\text{argMaxNoisyScore}]$
  
  Sample $\qtildewem \sim \text{Unif}\left[\text{left} , \text{right} \right]$
  
  \Return $\qtildewem$

    \caption{Widened Exponential Mechanism for Quantile ($\wexpm$): $\eps$-DP Algorithm}  \label{alg:wem}
\end{algorithm}

Below, we state the standard utility theorem for $\wexpm$ with widening hyperparameter $\theta$ on a fixed dataset $\slopes$. Let $\fsinv$ be the inverse empirical distribution function for the set of slopes, $\slopes$.  
We use the following assumption on the range and widening hyperparameters $R, \theta \in \posreals$. 
\begin{assumption} 
\label{assumption:range}
For a target $q \in (0,1)$ and fixed dataset $\slopes \in \reals^N$, the true quantile $\fsinv(q) \in [- R + \theta, R - \theta]$.
\end{assumption}

\begin{mytheorem}[Utility of \wexpm]
\label{thm:utility-of-wem}
Let $\slopes \in \reals^\numpairs$ be a fixed sample of $\numpairs$ points, and let $\fsinv$ be the inverse empirical distribution function. For $q \in (0,1)$, let $[-R,R]$, the range hyperparameter, satisfy Assumption~\ref{assumption:range}, and let $\theta, \eps > 0$. Let $\qtildewem = \texttt{DPWide}(\slopes, \eps, (q, -R, R, \theta)$.
Then, for $0 < c < \min(q, 1/2)$, 
\begin{align*}
\Pr_{ \substack{\wexpm(\slopes, \eps, \\ (q, -R, R, \theta))}}
\Bigl(\qtildewem \notin [\fsinv(q - c) - \theta, \fsinv(q + c) + \theta] \Bigr)
\leq \frac{R}{\theta} \exp \left(-\frac{\varepsilon c\numpairs}{2}\right)
\end{align*}
\end{mytheorem}
\begin{proof}
Let $\fs$ be the empirical distribution function for $\slopes$. The utility score of an output $r \in [-R, R]$ is 
$- \lfloor \min_{w \in [r-\theta, r+\theta]}
\numpairs \cdot \left \vert \fs(w) - q
\right\vert \rfloor$.
Let $r^* \in [\max(-R, r-\theta), \min(R, r+\theta)]$ be the value that maximizes the utility function for potential output value $r$.
Therefore, we can rewrite the utility score of $r$ as $- \lfloor \numpairs\left\vert \fs(r^*) - q \right\vert \rfloor$. 

Next, let us upper bound the probability that the mechanism selects an output with score $\leq -c\numpairs$.
The exponential mechanism assigns un-normalized probability density of at most $\exp(-\eps c\numpairs / 2)$ to each of these outputs, and they span at most the interval $[-R, R]$. 
On the other hand, the exponential mechanism assigns un-normalized probability density of $1$ to output values with score of zero, which we know exist in the range $[-R,R]$ by Assumption~\ref{assumption:range}. In particular, all outputs within $\theta$ of $\fsinv(q)$ have score of zero. Therefore, we have that
\begin{align*}
\Pr_{\wexpm(\bs, \eps, (q, -R, R, \theta))} &\Bigl( \qtildewemstar \notin [\fsinv(q - c), \fsinv(q + c)]\Bigr)  \\
    &= \Pr_{\wexpm(\slopes, \eps, (q, -R, R, \theta))} \Bigl( \left\vert\fs(\qtildewemstar) - q \right\vert \geq c \Bigr) \\
    &= \Pr_{\wexpm(\slopes, \eps, (q, -R, R, \theta))} \Bigl( \numpairs \left\vert\fs(\qtildewemstar) - \fs(\hat{q})\right\vert \geq c\numpairs \Bigr) \\
    &\leq 
    \frac{2R \exp(-\eps c\numpairs /2)}{2 \theta \exp(- \eps \cdot 0 \cdot \numpairs/2)} \\
    &= 
    \frac{R \exp(-\eps c\numpairs /2)}{\theta}
\end{align*}
Since $\qtildewem$ is at most $\theta$ away from $\qtildewemstar$, we can expand the interval by $\theta$ on each side and obtain the desired bound for $\qtildewem$.
\end{proof}
The widening parameter $\theta$ needs to be carefully chosen. All outputs within $\theta$ of the target quantile are given the same utility score, so a large $\theta$ represents a lower bound on the performance. Conversely, choosing $\theta$ too small may result in the area around the target quantile not being given sufficient weight in the sampled distribution. We describe how to set $\theta$ for \texttt{DPWideTS} in Section~\ref{sec:choosing-theta}.

\section{Finite-Sample Convergence of U-Statistic}

In this section, we develop a finite-sample convergence bound for U-statistics (Definition~\ref{def:u-statistic}) that will be a key component towards proving the finite-sample convergence bound for \texttt{DPWideTS}. The bound is stated in Theorem~\ref{thm:u-statistic-distribution-berry-esseen}.
To develop this bound, we adapt the convergence bounds for linear statistics of i.i.d. mean-0 random variables~\citep{chen2011normal} to work for U-statistics of independent, \emph{non-identical} random variables. 

We will start by pointing out another fact about U-statistics.
 \begin{fact}[\cite{Sen68}]
 \label{fact:u-nonincreasing}
 For all $z \in \reals$, $U_z$ is a non-increasing function.
 \end{fact}
Next, we analyze the expectation and variance of the U-statistic in question.
 
\subsection{Expectation and Variance of U-statistic}
\label{sec:expectation-variance}

Here, we consider the expectation and variance of $U_z$ (Equation \ref{eq:u-statistic-z-e}) in a very general form.
These expressions are not fully evaluated in order to be as general as possible. They can be further characterized (as shown in Section~\ref{sec:special-case-supplement}) based on features of the $x$ values (number of ties, spacings, etc.) and the paired datapoints that are included within the set $\matchedindexset$.
\begin{mylemma}
\label{lem:u-bn-statistic-expectation}
Let $x_1, \ldots, x_n$ satisfy Assumption~\ref{assumption:x-constants} and have variance $\sigma_x^2$, and let $y_1, \ldots, y_n$ be the corresponding response variables under the model $y_i = \intercept + \slope x_i + e_i$, where $\intercept, \slope \in \reals$ and each $e_i$ is sampled i.i.d from a continuous, symmetric, mean-0 distribution $F_e$ with variance $\sigma_e^2$. 
 Let $\matchedindexset$ be a set of unordered pairs of datapoints, let $\numpairs = |\matchedindexset|$, and let $U_z$ be defined accordingly as in Definition~\ref{def:u-statistic}, Equation (\ref{eq:u-statistic-z-e}). Let $\fdiff$ refer to the CDF of the difference in any two i.i.d. noise variables $e_j, e_i$, and let $\Delta_{ij} = x_j - x_i$. 
Then, we have that
\[
    \mathrm{E}_{\es}[U_z] = \frac{1}{\numpairsdenom} \sum_{\pairsum} \sign(x_j - x_i) \cdot \left( 1 - 2 \cdot \fdiff(z \cdot \Delta_{ij})\right)
\]
\end{mylemma}
\begin{proof}
Using the definition of $\fdiff$, we can expand the expectation of $U_z$ as follows. 
\begin{align*}
    \mathrm{E}_{\es}[U_z] 
    &= \frac{1}{\numpairsdenom} \sum_{\pairsum}\E_{e_i, e_j} \left[\sign(x_j - x_i) \cdot \sign \left(e_j - e_i - z \cdot \Delta_{ij} \right) \right] \\
    &= \frac{1}{\numpairsdenom} \sum_{\pairsum} \sign(x_j - x_i) \cdot \left( 1 - 2 \cdot \pr_{e_j, e_i}\left( e_j - e_i < z \cdot \Delta_{ij} \right)\right)\\
    &= \frac{1}{\numpairsdenom} \sum_{\pairsum} \sign(x_j - x_i) \cdot \left( 1 - 2 \cdot \fdiff(z \cdot \Delta_{ij})\right)
\end{align*}
\end{proof}


The next lemma bounds the variance of the U-statistic, $U_z$.
\begin{mylemma}
\label{lem:u-bn-statistic-variance}
Let $x_1, \ldots, x_n$ satisfy Assumption~\ref{assumption:x-constants} and have variance $\sigma_x^2$, and let $y_1, \ldots, y_n$ be the corresponding response variables under the model $y_i = \intercept + \slope x_i + e_i$, where $\intercept, \slope \in \reals$ and each $e_i$ is sampled i.i.d from a continuous, symmetric, mean-0 distribution $F_e$ with variance $\sigma_e^2$.  Let $\matchedindexset$ be a set of unordered pairs of datapoints, let $\numpairs = |\matchedindexset|$, and let $U_z$ be defined accordingly as in Definition~\ref{def:u-statistic}, Equation (\ref{eq:u-statistic-z-e}).
Let $B_{ij}(z) = \sign(x_j - x_i) \cdot \sign (e_j - e_i - z \cdot (x_j - x_i))$.
Then, we have that
\begin{align*}
    \frac{1}{\numpairsdenom^2} \sum_{\substack{\pairsum \\ \pairsumst \\ \pairsumoverlap{1}}} \Cov \left(B_{ij}(z), B_{st}(z) \right) \leq \Var_{\es}\left[U_z\right] 
    \leq
      \frac{1}{\numpairsdenom} + \frac{1}{\numpairsdenom^2} \sum_{\substack{\pairsum \\ \pairsumst \\ \pairsumoverlap{1}}} \Cov \left(B_{ij}(z), B_{st}(z) \right) 
\end{align*}
\end{mylemma}
\begin{proof}
We begin by rewriting the variance of $U_z$ as a sum of covariances. Then, we can split up the sum based on the number of overlaps of datapoints $i, j, s, t$ where $\{i, j\}, \{s, t\} \in \matchedindexset$.
\begin{align*}
    &\Var_{\es}\left[U_z\right] \\
    &= \frac{1}{\numpairsdenom^2} \sum_{\substack{\pairsum \\ \pairsumst} } \Cov \left( B_{ij}(z) \cdot B_{st}(z) \right) \\
    &= \frac{1}{\numpairsdenom^2} \left( \sum_{\pairsum} \Var \left(B_{ij}(z) \right)
    +
     \sum_{\substack{\pairsum \\ \pairsumst \\ \pairsumoverlap{1}}} \Cov \left(B_{ij}(z), B_{st}(z) \right) 
     + \sum_{\substack{\pairsum \\ \pairsumst \\ \pairsumoverlap{0}}} \Cov \left(B_{ij}(z), B_{st}(z) \right) \right) \\
     &\leq \frac{1}{\numpairsdenom} + 
     \left( \frac{1}{(\numpairsdenom)^2} \sum_{\substack{\pairsum \\ \pairsumst \\ \pairsumoverlap{1}}} \Cov \left(B_{ij}(z), B_{st}(z) \right) \right) + 0 
\end{align*}
The first sum in the second line can be evaluated by noting that $|\matchedindexset| = N$ and since $\Pr_{e_i, e_j}[B_{ij} \in \{-1, 1\}] = 1$, $\Var(B_{ij}(z)) \leq 1$. 
For the third sum, note that for $|\{i,j\} \cap \{s, t\}| = 0$, $B_{ij}(z)$ and $B_{st}(z)$ are independent, so $\Cov(B_{ij}(z), B_{st}(z)) = 0$. This leaves the second sum, which can be evaluated based on additional assumptions about the number of ties and spacings between $x$ values, as in Lemma~\ref{lem:u-endpoints-variance}.
Note that since $\Var(B_{ij}(z)) \geq 0$, we also have that
\begin{align*}
     Var_{\es}\left[U_z\right] \geq \frac{1}{\numpairsdenom^2} \sum_{\substack{\pairsum \\ \pairsumst \\ \pairsumoverlap{1}}} \Cov \left(B_{ij}(z), B_{st}(z) \right) 
\end{align*}
which gives the desired result.
\end{proof}

Using Lemma~\ref{lem:u-bn-statistic-variance}, we will evaluate the variance of the null U-statistic, $U_0$.
\begin{mycorollary}
\label{lem:u-null-statistic-variance}
Let $x_1, \ldots, x_n$ satisfy Assumption~\ref{assumption:x-constants} and have variance $\sigma_x^2$, and let $y_1, \ldots, y_n$ be the corresponding response variables under the model $y_i = \intercept + \slope x_i + e_i$, where $\intercept, \slope \in \reals$ and each $e_i$ is sampled i.i.d from a continuous, symmetric, mean-0 distribution $F_e$ with variance $\sigma_e^2$.  Let $\matchedindexset$ be a set of unordered pairs of datapoints, let $\numpairs = |\matchedindexset|$, and let $U_0$ be defined accordingly as in Definition~\ref{def:u-statistic}, Equation (\ref{eq:u-null-statistic}).  
Then, we have that
\begin{align*}
\Var_{\es}[U_0] 
    =
     \frac{1}{\numpairsdenom} 
     + \left( \frac{1}{3 \cdot \numpairsdenom^2} \sum_{\substack{\pairsum \\ \pairsumst \\ \pairsumoverlap{1}}} \sign(x_j - x_i) \cdot \sign (x_t - x_s) \right) 
\end{align*}
\end{mycorollary}
\begin{proof}
For any $\pairsum$, let $B_{ij}(0) = \sign(x_j - x_i) \cdot \sign (e_j - e_i)$. First, note that $\Var\left[B_{ij}(0)\right] = 1$ and $\Cov(B_{ij}(0), B_{st}(0))$ can be expanded as follows.
\begin{align*}
    &\Cov \left(B_{ij}(0), B_{st}(0) \right) \\
    &= \E_{e_i, e_j, e_s, e_t}[B_{ij}(0) \cdot B_{st}(0)] - \E_{e_i, e_j}[B_{ij}(0)] \cdot \E_{e_s, e_t} [B_{st}(0)] \\
    &= \sign(x_j - x_i) \cdot \sign (x_t - x_s) \cdot \left(  \E \left[\sign(e_j - e_i) \cdot \sign(e_t - e_s) \right] - \E \left[\sign(e_j - e_i) \right] \cdot \E \left[\sign(e_t - e_s)\right]  \right)
\end{align*}
It can be shown that for all $\pairsum$ and $\pairsumst$ such that $\pairsumoverlap{1}$, we have that $\E[\sign(e_j - e_i) \cdot \sign(e_t - e_s)] = 1/3$~\citep{Kendall48}.
In addition, by the symmetric, mean-0 nature of $F_e$, we have that $\E[\sign(e_j-e_i)] = \E[\sign(e_t - e_s)] = 0$.
Plugging these into the characterization of $\Var[U_z]$ from Lemma~\ref{lem:u-bn-statistic-variance} gives the desired result.
\begin{align*}
    \Var_{\es}[U_0] 
    &= \frac{1}{\numpairsdenom^2} \left( \sum_{\pairsum} \Var \left(B_{ij}(0) \right)  \right) +
     \left( \frac{1}{\numpairsdenom^2} \sum_{\substack{\pairsum \\ \pairsumst \\ \pairsumoverlap{1}}} \Cov \left(B_{ij}(0), B_{st}(0) \right) \right)  \\
    &=
     \frac{1}{\numpairsdenom} 
     + \left( \frac{1}{3 \cdot \numpairsdenom^2} \sum_{\substack{\pairsum \\ \pairsumst \\ \pairsumoverlap{1}}} \sign(x_j - x_i) \cdot \sign (x_t - x_s) \right) 
\end{align*}
\end{proof}

\subsection{Berry-Esseen-type bound for U-statistic}

Now, we move onto developing the finite-sample convergence bound for independent, non-identical U-statistics, shown in Theorem~\ref{thm:u-statistic-distribution-berry-esseen}.
First, we state a standard Berry-Esseen theorem for independent, mean-0, (not necessarily identical) statistics.
\begin{mytheorem}[Berry-Esseen bound for independent mean-0 r.v.s~\citep{berry1941accuracy,esseen1942liapunov}]
\label{thm:berry-esseen-non-iid}
Let $\zeta_1, \ldots, \zeta_n$ be independent random variables with $\mathrm{E}[\zeta_i] = 0$, 
$\mathrm{E}[\zeta_i^2]  > 0$, 
and $\mathrm{E}[|\zeta_i|^3] = \rho_i^3 < \infty$, for $i \in [n]$. Let $\sum_{i=1}^n \Var(\zeta_i) = 1$ and $\rho^3 = \sum_{i=1}^n \rho_i^3$. 
Let $W = \sum_{i=1}^n \zeta_n$ and $\Phi$ be the standard normal cdf. Then, 
\begin{align*}
    \sup_{t \in \reals} |\Pr(W \leq t) - \Phi(t)| \leq C \cdot \rho^3
\end{align*}
where $C>0$ is a universal constant.
\end{mytheorem}

Now, to prove Theorem~\ref{thm:u-statistic-distribution-berry-esseen}, we begin by stating an inequality presented by~\cite{chen2011normal} that bounds the difference between the cumulative distributions of a non-linear function $T$ and a linear approximation function $W$. Note that this inequality, stated below in Lemma~\ref{lem:randomized-concentration-inequality}, does not require i.i.d random variables, but the subsequent theorems in~\cite{chen2011normal} that rely on this bound do require identical marginals.

\begin{mylemma}[\cite{chen2011normal}]
\label{lem:randomized-concentration-inequality}
Let $\zeta_1, \ldots, \zeta_n$ be independent random variables satisfying $\mathrm{E}[\zeta_i] = 0$, $\mathrm{E}[\zeta_i^2] > 0$, and $\mathrm{E}[|\zeta_i|^3] < \infty$ for $i \in [n]$. Let $\sum_{i=1}^n \Var(\zeta_i) = 1$, $W = \sum_{i=1}^n \zeta_i$, and $T = W+\Delta$ for some $\Delta := \Delta(\zeta_1, \ldots, \zeta_n)$. For each $i \in [n]$, let $\Delta_i$ be a random variable such that $\zeta_i$ and $(W-\zeta_i, \Delta_i)$ are independent.
Then, for all $t \in \reals$,
\begin{align*}
    \big \vert \Pr(T \leq t) - \pr(W \leq t) \big \vert 
    \leq 
    2 \sum_{i=1}^n \E[\zeta_i^2] \ind_{|\zeta_i|>1} 
    +
    2 \sum_{i=1}^n \E[|\zeta_i|^3] \ind_{|\zeta_i|\leq 1} 
    +
    \E \left[ \left \vert W\Delta \right \vert\right] 
    + 
    \sum_{i=1}^n\E\left[ \left \vert \zeta_i (\Delta - \Delta_i) \right \vert \right].
\end{align*}
\end{mylemma}
Using the inequality above in Lemma~\ref{lem:randomized-concentration-inequality}, along with a standard Berry-Esseen theorem (Theorem~\ref{thm:berry-esseen-non-iid}) to replace $\Pr[W \leq t]$ with $\Phi(t)$,~\cite{chen2011normal} develop a Berry-Esseen bound for non-linear statistics of i.i.d. mean-0 random variables. We will adapt this bound for non-identical yet independent, mean-0 random variables to match the setting of our U-statistic, $U_z$.

To do so, we first state some definitions that allow us to rewrite $U_z$ as the sum of a linear approximation function $W$ and a remainder $\Delta$. These include defining the variables $\Psi_1, \ldots, \Psi_n$ and $\zeta_1, \ldots, \zeta_n$. To help parse the notation in the following lemma, note that each $\Psi_i$ is conditioned on the corresponding noise variable $e_i$, so
$\Psi_1, \ldots, \Psi_n$ are independent. In addition, note that the expectation $\denommu$ and variance $\denomvar$ of the sum of the $\Psi_i$'s can be related to the expectation and variance of $U_z$ (as shown in Lemma~\ref{lem:denom-var}). The linear approximation function $W$ is the sum of $\zeta_i$, $i \in [n]$, which are normalized versions of the $\Psi_i$ random variables.

\begin{mylemma}
\label{lem:u-statistic-linear-approximation}
Let $\matchedindexset$ be a set of unordered pairs of datapoints from $\datatuples$, where $|\matchedindexset| = \numpairs = nk/2$
. For $\pairsum$, let $a_{ij} = \sign(x_j - x_i)$, $b_{ij} = \sign(e_j - e_i)$, and
$c_{ij}(z) = \sign\left(e_j - e_i - z \cdot (x_j - x_i)\right)$, and $B'_{ij} = \E_{e'}\left[\sign\left(e' - e_i - z \cdot (x_j - x_i)\right) \mid e_i \right]$ where $e'$ is a fresh draw from the distribution $F_e$.
Then, for $i \in [n]$, let 
\[
\Psi_i =  \frac{1}{k} \sum_{j: \pairsumdouble} a_{ij} \cdot B'_{ij} ~~~\text{and}~~~ \zeta_i = \frac{\Psi_i - \mu_i}{\denomvar \cdot n}
\]
where $\mu_i = \E_{\es}[\Psi_i]$ and
$\sigma_i^2 = \Var_{\es}[\Psi_i]$. In addition, let  $\mu_{\Psi} = \sum_{i=1}^n \mu_i / n$ and $\sigma_\Psi^2 = \sum_{i=1}^n \sigma_i^2 / n^2$. 
Now, we define the following:
\begin{align*}
\bar{h}(i, j) &= a_{ij} \cdot c_{ij} - 
\frac{1}{2} \cdot \Psi_i - 
\frac{1}{2} \cdot \Psi_j \\
    \Delta &= \frac{1}{2 \cdot \denomvar \cdot \numpairs} \sum_{\pairsumdouble} \bar{h}(i,j) ~~~~~\text{and}~~~~~~
    \Delta_l = \frac{1}{2 \cdot \denomvar \cdot \numpairs} \sum_{\substack{\pairsumdouble \\ i, j \neq l}} \bar{h}(i,j)
\end{align*}
Finally, let $W = \sum_{i=1}^n \zeta_i$.
Then, we claim the following.
\begin{enumerate}
\item $\zeta_1, \ldots, \zeta_n$ are independent
\item $W + \Delta = \left( U_z - \denommu \right) / (\denomvar)$ 
\item For each $l \in [n]$, $\zeta_l$ is independent of $(W - \zeta_l, \Delta_l)$. 
\end{enumerate}
\end{mylemma}
\begin{proof}
First, note that each $\zeta_i$ conditions on the corresponding noise variable $e_i$, and the only randomness in each $\zeta_i$ is a fresh draw $e'$ from the distribution $F_e$, so $\zeta_1, \ldots, \zeta_n$ are independent.

Next, using the definitions of $\Psi_i$ and $\bar{h}(i,j, z)$, we can expand $W +\Delta$ as follows. In particular, note that the $\Psi_i$ and $\Psi_j$ variables cancel out, so we are left with sums of $a_{ij}c_{ij}$ and $\mu_i$.
\begin{align*}
    W + \Delta
    &= \sum_{i=1}^n \zeta_i + \frac{1}{ \denomvar \cdot \numpairs} \sum_{\pairsumdouble} \bar{h}(i, j) \\
    &= \frac{1}{\denomvar \cdot n} \sum_{i=1}^n (\Psi_i - \mu_i) +
    \frac{1}{2 \cdot \denomvar \cdot \numpairs} \sum_{\pairsumdouble} \left(a_{ij} c_{ij} - \frac{1}{2} \cdot \Psi_i - \frac{1}{2} \cdot \Psi_j \right) \\
    &= \frac{1}{2 \cdot \denomvar \cdot \numpairs} \sum_{\pairsumdouble} a_{ij} c_{ij}  - \frac{1}{\denomvar \cdot n} \sum_{i=1}^n \mu_i \\
    &~~~~~~~+ \frac{1}{\denomvar \cdot n} \sum_{i=1}^n \Psi_i - \frac{1}{4 \cdot \denomvar \cdot \numpairs} \sum_{\pairsumdouble} \left( \Psi_i + \Psi_j \right)\\
    &= \frac{1}{\denomvar \cdot \numpairs} \sum_{\pairsum} a_{ij} c_{ij}  - \frac{1}{\denomvar \cdot n} \sum_{i=1}^n \mu_i \\
    &= \frac{U_z - \denommu}{\denomvar} 
\end{align*}
Finally, note that for each $l \in [n]$, the random variables $W - \zeta_l$ and $\Delta_l$ are functions of $\zeta_j$, $j \neq l$. As $\zeta_1, \ldots, \zeta_n$ are independent of each other, this means that $\zeta_l$ is independent of $(W - \zeta_l, \Delta_l)$.
\end{proof}

Below, we relate the quantities $\denommu$ and $\denomvar^2$ to $\mu(z)$ and $\sigma^2(z)$ , the expectation and variance of $U_z$.
\begin{mylemma}
\label{lem:denom-var}
Let $\matchedindexset$ be the set of unordered pairs of datapoints from $\datatuples$, 
and let $\numpairs = |S|$.
Let $\denommu, \denomvar$ be defined as in Lemma~\ref{lem:randomized-concentration-inequality}. In addition, let $U_z$ be defined as in (\ref{eq:u-statistic-z-e}), namely $\mu(z) = \E_{\es}[U_z]$, and $\sigma^2(z) = \Var_{\es}[U_z]$.
Then,
\begin{align*}
    \denommu = \mu(z)  ~~~\text{and}~~~
    \denomvar^2 = \sigma^2(z) + \Theta(1/N)
\end{align*}
\end{mylemma}
\begin{proof}
First, we show that $\denommu(z)$ can be rewritten as $\mu(z)$. Let $B_{ij} = E_{e'} \left[ e_j - e_i - z (x_j - x_i) ~\vert~ e_i \right]$ and $B'_{ij} = E_{e'} \left[ e' - e_i - z (x_j - x_i) ~\vert~ e_i \right]$, where $e'$ is a fresh draw of the random variable $e$. Note that as $e_j$ and $e'$ are identical random variables, both of which are independent from $e_i$, $B_{ij} = B'_{ij}$. Then, we have that
\begin{align*}
    \denommu 
    &= \frac{1}{n} \sum_{i=1}^n \E_{\es}\left[ \Psi_i \right] \\
    &= \frac{1}{n} \sum_{i=1}^n \E_{\es}\left[ \frac{1}{k} \sum_{j: \pairsumdouble} a_{ij} \cdot B'_{ij} \right] \\
    &= \frac{2}{nk} \sum_{\pairsum} a_{ij} \cdot B_{ij} \\
    &= E_{\es}\left[ U_z \right] \\
    &= \mu(z)
\end{align*}
Next, for $\sigma(z)$, we know from Lemma~\ref{lem:u-bn-statistic-variance} that
\begin{align*}
\frac{1}{\numpairsdenom^2} \sum_{\substack{\pairsum \\ \pairsumst \\ \pairsumoverlap{1}}} \Cov \left(B_{ij}(z), B_{st}(z) \right) 
    \leq \sigma^2(z) 
     \leq \frac{1}{\numpairsdenom} + 
     \frac{1}{\numpairsdenom^2} \sum_{\substack{\pairsum \\ \pairsumst \\ \pairsumoverlap{1}}} \Cov \left(B_{ij}(z), B_{st}(z) \right)  
\end{align*}
In addition, note that $\Cov(B_{ij}', B_{st}'') = \Cov(B_{ij}, B_{st})$. To see this, notice that $e'$ is identical to $e_j$, and both are independent from the other random variables $e_i, e_s, e_t, e''$, so $e',e_j$ are interchangeable in this expression. Similarly, $e''$ is identical to $e_t$, and both are independent from $e_i, e_j, e', e_s$, so $e''$ and $e_t$ are interchangeable in this expression.

Then, note that we can rewrite $\denomvar$ as follows.
\begin{align*}
    \denomvar^2 
    &= \frac{1}{n^2} \sum_{i=1}^n \Var[\Psi_i] \\
    &= \frac{1}{n^2} \sum_{i=1}^n \left( \frac{1}{k^2} \sum_{\substack{j: \pairsumdouble \\ t: \{i, t\} \in \matchedindexset}} \Cov(B_{ij}', B_{it}'') \right) \\
    &= \frac{1}{n^2 k^2} \sum_{i=1}^n \left( \sum_{j: \pairsumdouble} \Cov(B_{ij}', B_{ij}'') + \sum_{\substack{j: \pairsumdouble \\ t: \{i, t\} \in \matchedindexset, t \neq j}} \Cov(B_{ij}', B_{it}'') \right) \\
     &= \frac{2}{n^2 k^2} \sum_{ \pairsum} \Cov(B_{ij}', B_{ij}'') + \frac{4}{n^2 k^2} \sum_{\substack{\pairsum \\ \pairsumst \\ \pairsumoverlap{1} }} \Cov(B_{ij}', B_{st}'') \\
    &= \frac{1}{N} \sum_{\pairsum} \Cov(B_{ij}', B_{ij}'') + \frac{1}{N^2} \cdot \sum_{\substack{\pairsum \\ \{s, t\} \in \matchedindexset \\ |\{i, j\} \cap \{s, t \}| = 1}} \Cov(B_{ij}, B_{st}) \\
    &= \sigma^2(z) + \Theta \left(\frac{1}{N}\right) 
\end{align*}
As we can see, this gives the desired result. 
\end{proof}

Next, we use bounds for $\E[\Delta^2]$ and $\E[(\Delta-\Delta_l)^2]$ shown by~\cite{chen2011normal}. The proof is lengthy so it is omitted here.


\begin{mylemma}[\cite{chen2011normal}]
\label{lem:delta-l}
Let $\Delta, \Delta_l$ and $\zeta_l$ (for $l \in [n]$) be defined as in Lemma~\ref{lem:u-statistic-linear-approximation}. In addition, for $U_z$ as defined in (\ref{eq:u-statistic-z-e}), let $\sigma^2(z) = \Var_{\es}[U_z]$, and let $\denomvar^2 = \sum_{l=1}^n \Var[\Psi_l] / n^2$. Then, 
\begin{align*}
\E\left[\Delta^2\right] \leq  \frac{\sigma^2(z)}{2 (n-1) \cdot \denomvar^2}  ~~~\text{and}~~~
\E\left[(\Delta - \Delta_l)^2\right] \leq \frac{2\sigma^2(z)}{2n(n-1) \cdot \denomvar^2} 
\end{align*}
\end{mylemma}

The above lemma allows us to bound the right side of the inequality in Lemma~\ref{lem:randomized-concentration-inequality}.

\begin{mylemma}
\label{lem:randomized-concentration-implementation}
Let $W, \Delta, \Delta_l$, and $\zeta_l$ (for $l \in [n]$) be defined as in Lemma~\ref{lem:u-statistic-linear-approximation}, where $\rho^3 = \sum_{l=1}^n \E[|\zeta_l|^3]$. Let $\gamma$ be defined as follows.
\begin{align*}
    \gamma =
    C \cdot \rho^3 
    + 
    2 \sum_{i=1}^n \E[\zeta_i^2] \ind_{|\zeta_i|>1} 
    +
    2 \sum_{i=1}^n \E[|\zeta_i|^3] \ind_{|\zeta_i|\leq 1} 
    +
    \mathrm{E} \left[ \left|W\Delta\right|\right] + \sum_{l=1}^n \mathrm{E}\left[\left|\zeta_l(\Delta - \Delta_l)\right|\right]
\end{align*}
where $C > 0$ is a universal constant. . In addition, let $\sigma^2(z) = \Var_{\es}[U_z]$, where $U_z$ is defined in (\ref{eq:u-statistic-z-e}). Then, for sufficiently large $n$, $\gamma = O \left( 1 / (n^2 \sigma^3(z)) \right)$.
\end{mylemma}
\begin{proof}
\label{proof:randomized-concentration-implementation}
First, note that $|\Psi_i - \mu_i| \leq 2$ for all $i \in [n]$, because this term consists of a product and expectation of signs. Therefore, we have that 
\[
\left \vert \zeta_i \right \vert 
= \left \vert \frac{\Psi_i - \mu_i}{2 \cdot \denomvar \cdot n} \right \vert 
\leq \frac{1}{\denomvar \cdot n}, 
~~~ \rho^3 
= \sum_{i=1}^n \mathrm{E} \left[ \left \vert \zeta_i \right \vert^3 \right] 
\leq \frac{1}{\denomvar^3 \cdot n^2} 
\]
Note that $\denomvar = \Theta(1/\sqrt{n})$, so $\zeta_i = \Theta(1/\sqrt{n})$. Thus, for sufficiently large $n$, we have that
\begin{align*}
    \sum_{i=1}^n \E[\zeta_i^2] \cdot \ind_{|\zeta_i|>1} 
    +
    2 \sum_{i=1}^n \E[|\zeta_i|^3] \cdot \ind_{|\zeta_i|\leq 1} 
    = 0 + \rho^3 \leq \frac{1}{\denomvar^3 \cdot n^2}
\end{align*}
Next, note that $\E[W] = \E\left[\sum_{i=1}^n \zeta_i \right] = 0$. Therefore,
\begin{align*}
    \E[W^2] = \Var[W] = \frac{1}{4 \cdot \denomvar^2 \cdot n^2} \sum_{i \in [n]} \sigma_i^2 
     = O\left( \frac{n^2 \denomvar^2}{\denomvar^2 n^2} \right) = O(1)
\end{align*}
Using this and the fact from Lemma~\ref{lem:delta-l} that $E[\Delta^2] = O \left(\sigma^2(z) / n \cdot \denomvar^2 \right)$, we have that
\begin{align*}
    \E[|W\Delta|]
    &\leq \E[|W||\Delta|] \\
    &\leq \sqrt{\E\left[|W|^2\right] \cdot \E \left[|\Delta|^2\right]} \\
    &= \sqrt{\E\left[W^2\right] \cdot \E \left[\Delta^2\right]} \\
    &= \sqrt{\E\left[W^2\right] \cdot \E \left[\Delta^2\right]} \\
    &= O \left( \frac{\sigma(z)}{ \sqrt{n} \cdot \denomvar} \right)
\end{align*}
Similarly, using the fact from Lemma~\ref{lem:delta-l} that $E[(\Delta-\Delta_l)^2] = O(\sigma^2(z) / n^2 \cdot \denomvar^2)$, we have that
\begin{align*}
    \sum_{l\in[n]} \E \left[ | \zeta_l \cdot (\Delta - \Delta_l) | \right] 
    &\leq 
    \sum_{l \in [n]} \sqrt{ \E\left[ \zeta_l^2 \right] \cdot \E\left[ (\Delta-\Delta_l)^2 \right] } \\
    &= \sum_{l \in [n]} \sqrt{ \frac{\sigma_l^2}{n^2 \cdot \denomvar^2} \cdot O\left( \frac{\sigma^2(z)}{n^2 \cdot \denomvar^2} \right) } \\
    &= O \left( \frac{\sigma(z)}{n^2 \cdot \denomvar^2} \right) \cdot \sum_{l\in[n]} \sigma_l \\
    &= O \left( \frac{\sigma(z)}{n^2 \cdot \denomvar^2} \right) \cdot O(n) \\
    &= O\left( \frac{\sigma(z)}{n \cdot \denomvar^2} \right)
\end{align*}
where the second to last line follows from noting that $\sigma_l \in (0, 1]$.
This gives us that
\begin{align*}
 \gamma 
 = O \left( \frac{1}{n^2 \cdot \denomvar^3} \right) + O\left(\frac{\sigma(z)}{ \sqrt{n} \cdot \denomvar}\right) + O\left(\frac{\sigma(z)}{n \cdot \denomvar^2} \right)
 = O\left( \frac{1}{n^2 \cdot \sigma^3(z)} \right)
\end{align*}
where the second step follows from noting that $\denomvar < 1$ for sufficiently large $n$, and by replacing $\denomvar$ with $\sigma(z)$ via Lemma~\ref{lem:denom-var}. 
\end{proof}

Below, we put all the above lemmas together to state and prove the finite-sample convergence of the distribution of the U-statistic to the standard normal distribution.

\begin{thmn}[~\ref{thm:u-statistic-distribution-berry-esseen}]
Let $x_1, \ldots, x_n$ satisfy Assumption~\ref{assumption:x-constants} (ie. not all equal) and let $y_1, \ldots, y_n$ be the corresponding response variables under the model $y_i = \intercept + \slope x_i + e_i$, where $\intercept, \slope \in \reals$ and each $e_i$ is sampled i.i.d from a continuous, symmetric, mean-0 distribution. 
Let $\matchedindexset$ be a set of unordered pairs of datapoints, 
 and 
let $U_z$ be defined with respect to $\matchedindexset$ as in (\ref{eq:u-statistic-z-e}). 
Let $\mu(z) = \E_{\es}[U_z]$, and $\sigma^2(z) = \Var_{\es}[U_z]$.
Then, for sufficiently large $n$,
\[
    \sup_{t \in \reals} \left \vert \Pr \left[  \frac{U_z - \mu(z)}{\sigma(z)} \leq t \right] - \Phi(t)
    \right \vert
    = O \left( \frac{1}{n^2 \cdot \sigma^3(z)} \right) 
\]
where $\Phi$ is the cdf of a standard normal distribution.
\end{thmn}
\begin{proof}
\label{proof:u-statistic-distribution-berry-esseen}
Recall from Lemma~\ref{lem:u-statistic-linear-approximation} the following definitions: for $a_{ij} = \sign(x_j - x_i)$ and \\
$B_{ij}' = \E_{e'} \left[ \sign\left(e' - e_i - z \cdot (x_j - x_i)\right) \mid e_i \right]$, where $e'$ is a fresh draw of the random variable $e$, we let
\begin{align*}
    \Psi_i = \frac{1}{k} \sum_{j: \pairsumdouble} a_{ij} \cdot B_{ij}'
    ~~\text{and}~~
        \zeta_i = \frac{\Psi_i - \mu_i}{\denomvar \cdot n}
\end{align*}
where $\mu_i = \E_{\es}[\Psi_i]$, $\sigma_i = \Var_{\es}[\Psi_i]$, and $\denomvar^2 = \sum_{i=1}^n \sigma_i^2/n^2$. From these definitions, it can be shown that $E[\zeta_i] = 0, E[\zeta_i^2] > 0, E[|\zeta_i|^3] < \infty$, and $\sum_{i=1}^n \Var[\zeta_i] = 1$. In addition, Lemma~\ref{lem:denom-var} shows that for $\denommu = \sum_{i=1}^n \mu_i/ n$,
\begin{align*}
    \mu(z) = \denommu, ~~~~ \sigma^2(z) = \denomvar^2 + \Theta(1/\numpairs),
\end{align*}
where $\numpairs = |\matchedindexset|$. By Lemma~\ref{lem:u-statistic-linear-approximation}, we therefore have that
\begin{align*}
    \frac{U_z - \mu(z)}{\sigma(z)} = W+\Delta,
\end{align*}
and for each $i \in [n]$, $\zeta_i$ is independent of $(W - \zeta_i, \Delta_i)$. Then, putting together Lemma~\ref{lem:randomized-concentration-inequality} and Lemma~\ref{lem:randomized-concentration-implementation} directly gives the desired result. 
\end{proof}

\section{Finite-Sample Convergence Bound for \texttt{DPWideTS}}
\label{sec:wemts-proof}

Here, we prove our main result: the finite-sample convergence bound for $\betatildewemts$. 
As described earlier in Section~\ref{sec:overview-of-analysis}, the proof of Theorem~\ref{thm:wemts-bound} follows in two main steps; first, we argue that $U(\betatildewemts)$ is close to 0 with high probability by utility theorem of the exponential mechanism (Theorem~\ref{thm:utility-of-wem}). Second, we use the convergence of the U-statistic distribution (Theorem~\ref{thm:u-statistic-distribution-berry-esseen}) to show that $\betatildewemts$ is close to $\slope$.

\begin{thmn}[\ref{thm:wemts-bound}]
Let $x_1, \ldots, x_n$ satisfy Assumption~\ref{assumption:x-constants} and have empirical variance $\sigma_x^2$. Let $y_1, \ldots, y_n$ be the corresponding response variables under the model $y_i = \intercept + \slope x_i + e_i$, where $\intercept, \slope \in \reals$ and each $e_i$ is sampled i.i.d from a continuous, symmetric, mean-0 distribution $F_e$. 
Let $\betatildewemts = \texttt{DPTheilSen}(\{x_i, y_i\}_{i=1}^n, \eps, (\matchedindexset, \wexpm, \theta, -R, R))$, where $\eps, R, \theta >0$, $\slope \in [-R+\theta, R-\theta]$\removeforsubmission{ as in Assumption~\ref{assumption:range}}, and $\matchedindexset \in \binom{[n]}{2}$ is a set of unordered pairs with distinct x-values. Let $U_z$ be the U-statistic defined with respect to $\matchedindexset$.

Then, there exists a constant $c >0$ such that for sufficiently large $n$, with probability at least $1-p$, $\betatildewemts \in [ \slope - z - \theta, \slope + z + \theta]$ for every $z$ that satisfies the following:
\begin{align*}
    \frac{-\mu(z) - \frac{4\ln(4R/p\theta)}{\eps \cdot n}}{\sigma(z)} \geq \Phi^{-1}\left( 1 - \frac{p}{4} + \frac{c}{n^2 \cdot \sigma^3(z)} \right)
\end{align*}
where $\mu(z) = \E[U_z], \sigma^2(z) = \Var[U_z]$, and $\Phi$ is the standard normal cdf.
\end{thmn}

\begin{proof}
First, we consider the event that $\betatildewemts > \slope + z + \theta$. 
We take a union bound over two possibilities that could lead to this event: the first is that $U(\betatildewemts-\theta)$ is less than an arbitrary value $-c \in (-1, 0)$ (ie. $\wexpm$ returned an output that is more than $cN/2$ away in rank from the median of the $N$ slopes), which implies that $\betatildewemts$ is not within $z$ of $\slope$ with high probability. The second that $U(\betatildewemts-\theta)$ is within $c$ of $0$, yet $\betatildewemts-\theta$ is still more than $z$ greater than $\slope$. We can simplify these expressions by noting that $U$ is a non-increasing function (Fact~\ref{fact:u-nonincreasing}). 
Then, for all $z \in \reals$ and sufficiently large $n$, we have that
\begin{align*}
    \Pr_{Y, \wexpm}&\left[ \betatildewemts > \slope + z + \theta \right] \\
    &\leq \Pr_{Y, \wexpm} \left[ U(\betatildewemts - \theta, \datatuples) < -c \right] + \\
   &~~~~ \Pr_{Y, \wexpm} \Big[ U(\betatildewemts - \theta , \datatuples) \geq -c  \cap \betatildewemts - \theta > \slope + z \Big] \\
    &\leq \Pr_{Y, \wexpm} \left[ U(\betatildewemts, \datatuples) < -c \right] + \Pr_{Y, \wexpm} \left[ U(\slope + z + \theta, \datatuples) \geq -c \right] \\
    &\leq \Pr_{Y, \wexpm} \left[ 1/2 - \fs(\betatildewemts) < -c/2 \right] +  \Pr_{Y, \wexpm} \left[ U(\slope + z, \datatuples) \geq -c \right] \\
    &\leq \frac{R}{\theta} \exp\left( - \eps \cdot c \cdot N / 4 \right) + \left( 1 - \Phi\left( \frac{ - c - \mu(z)}{\sigma(z)} \right) \right) +  O\left(\frac{1}{n^2 \cdot \sigma^3(z) } \right)
\end{align*}
where $\fs$ is the empirical distribution function for the set of slopes computed by \texttt{DPWideTS}. Then, the first term in the last line comes from Theorem~\ref{thm:utility-of-wem} and the second term follows from Lemma~\ref{thm:u-statistic-distribution-berry-esseen}.
We can run through a similar analysis for the event that $\betatildewemts < \slope - z - \theta$ to get the following:
\begin{align*}
    \Pr_{Y, \wexpm}\left[ \betatildewemts < \slope - z - \theta \right] 
    \leq \frac{R}{\theta} \exp\left( - \eps \cdot c \cdot N / 4 \right) +  \Phi\left( \frac{ c - \mu(-z)}{\sigma(-z)} \right)  +
    O\left(\frac{1}{n^2 \cdot \sigma^3(-z) } \right)
\end{align*}
Setting the first terms less than or equal to $p/4$ and then solving for $c$ and $z$ gives the desired result. Finally, the expressions for $\mu(z)$ and $\sigma(z)$ follow directly from Lemmas~\ref{lem:u-bn-statistic-expectation} and~\ref{lem:u-bn-statistic-variance}.
\end{proof}

This result is not very easily interpretable in the general case. In the next section, we apply this bound to a special setting, and in Section~\ref{sec:special-case}, we provide intuition and comparison with convergence bounds for other non-DP and DP algorithms.

\section{Evaluating \texttt{DPWideTS} Bound for Special Case}
\label{sec:special-case-supplement}
We consider the special setting where the x-values are evenly split between the endpoints of an interval of length $\Delta_x$. The \texttt{DPWideTS} algorithm computes $N = \lfloor n/2 \rfloor \cdot \lceil n/2 \rceil$ slopes. This section evaluates the bound from Theorem~\ref{thm:wemts-bound} for this special setting.

To do so, we first solve for the expectation and variance of the U-statistic for this case so that we can plug these values into the general convergence bound.
We rely on the following approximation of the normal CDF.
\begin{mylemma}[Normal CDF approximation]
\label{lem:normal-cdf}
Let $F(\cdot)$ be the cumulative distribution function for a Gaussian with mean $\slope$ and variance $\sigma^2$. 
For all $y$ such that $|y-\slope|/\sigma$ is sufficiently small, we have that
\begin{align*}
    F(y) = \frac{1}{2} + \frac{y-\slope}{\sigma \sqrt{2 \pi}} + O\left( \frac{(y-\slope)^3}{\sigma^3} \right)
\end{align*}
\end{mylemma}
\begin{proof}
For any $y \in \reals$, the distribution function $F(y)$ is defined as follows.
\begin{align*}
    F(y) = \frac{1}{2}\left( 1 + \text{erf}\left(\frac{y - \slope}{ \sigma \sqrt{2} }\right)\right)
\end{align*}
The error function $\text{erf}(z)$ can be rewritten using a power series expansion. For $z$ close to $0$, we have:
\begin{align*}
    \text{erf}(z) 
    &= \frac{2}{\sqrt{\pi}} \int_{0}^{z} e^{-t^2} dt \\
    &= \frac{2}{\sqrt{\pi}} \int_{0}^{z} \left(1 - t^2 + \frac{t^4}{2!} + \ldots \right) dt \\
    &= \frac{2}{\sqrt{\pi}} \left(z - \frac{z^3}{3} + \ldots\right) \\
    &= \frac{2}{\sqrt{\pi}}z + O(z^3)
\end{align*}
Setting $z = (y-\slope)/ (\sigma \sqrt{2})$, we have that when $z$ is close to 0,
\begin{align*}
    F(y) 
    &= \frac{1}{2}\left( 1 + \frac{\sqrt{2}(y - \slope)}{ \sqrt{\pi} \sigma} + O\left( \frac{(y-\slope)^3}{\sigma^3} \right)\right)   \\
    &= \frac{1}{2} + \frac{y - \slope}{ \sigma \sqrt{2\pi}} + O\left( \frac{(y-\slope)^3}{\sigma^3} \right)
\end{align*}
which gives us the desired result.
\end{proof}

Now, we can evaluate the expectation and variance of $U_z$ for this special setting.
\begin{mylemma}
\label{lem:u-endpoints-expectation}
Suppose that $x_1, \ldots, x_n$ be at two endpoints of an interval of size $\Delta_x$ such that they satisfy Assumption~\ref{assumption:endpoints}, and let $e_1, \ldots, e_n$ be drawn i.i.d from $F_e = \calN(0, \sigma_e^2)$ according to Assumption~\ref{assumption:normal-errors}. Let $y_1, \ldots, y_n$ be the corresponding response variables under the model $y_i = \intercept + \slope x_i + e_i$, where $\intercept, \slope \in \reals$. Let $\matchedindexset$ be the set of unordered pairs of datapoints
, and let $U_z$ be defined accordingly as in (\ref{eq:u-statistic-z-e}).
Then, we have that for sufficiently small $z \cdot \Delta_x / \sigma_e$,
\begin{align*}
    \E_{\es} \left[U_z\right] = \frac{-z \cdot \Delta_x}{\sqrt{\pi} \sigma_e} +  O \left(\frac{z^3 \cdot \Delta_x^3}{\sigma_e^3} \right)
\end{align*}
\end{mylemma}
\begin{proof}
First, let $\fdiff$ refer to the CDF of the difference in any two i.i.d. noise variables $e_j, e_i$. In addition, let $N = |\matchedindexset|$. 
Then, we have from Lemma~\ref{lem:u-bn-statistic-expectation} that
\begin{align*}
    \mathrm{E}_{\es}[U_z] = \frac{1}{\numpairs} \sum_{\pairsum} \sign(\Delta_x) \cdot \left( 1 - 2 \cdot \fdiff(z \cdot \Delta_{x})\right)
\end{align*}
Next, note that for our special case (assumptions \ref{assumption:endpoints} and \ref{assumption:normal-errors}), $\sign(\Delta_x) = 1$ and $\fdiff$ is the CDF of the difference between two i.i.d noise variables from a $\calN(0, \sigma_e^2)$ distribution. Therefore, using a Taylor approximation of the normal cdf from Lemma~\ref{lem:normal-cdf}, we have that for sufficiently small $z \cdot \Delta_x / \sigma_e$,
\begin{align*}
    \E_{\es} \left[U_z \right] 
    &= \frac{1}{\numpairs} \sum_{\pairsum} \left(1 - 2 \cdot \fdiff (z \cdot \Delta_x)\right) \\ 
    &= \frac{1}{\numpairs} \sum_{\pairsum} \left( 1 - 2 \left(\frac{1}{2} + \frac{z \Delta_x}{2 \sqrt{\pi} \sigma_e} + O \left( \frac{z^3 \Delta_x^3}{\sigma_e^3} \right) \right) \right) \\
    &= \frac{-z \cdot \Delta_x}{\sqrt{\pi} \sigma_e} + O \left(\frac{z^3 \cdot \Delta_x^3}{\sigma_e^3} \right) 
\end{align*}
which gives the desired result.
\end{proof}

\begin{mylemma}
\label{lem:u-endpoints-variance}
Suppose that $x_1, \ldots, x_n$ be at two endpoints of an interval of size $\Delta_x$ such that they satisfy Assumption~\ref{assumption:endpoints}, and let $e_1, \ldots, e_n$ be drawn i.i.d from $F_e = \calN(0, \sigma_e^2)$ according to Assumption~\ref{assumption:normal-errors}.  Let $y_1, \ldots, y_n$ be the corresponding response variables under the model $y_i = \intercept + \slope x_i + e_i$, where $\intercept, \slope \in \reals$. Let $\matchedindexset$ be the set of unordered pairs of datapoints
, and let $U_z$ be defined accordingly as in (\ref{eq:u-statistic-z-e}).
Then, for sufficiently small $z \cdot \Delta_x / \sigma_e$,
\begin{align*}
    \Var_{\es} \left[U_z \right]
    &=
    \frac{4}{3n} + O\left( \frac{z \cdot \Delta_x}{n \cdot \sigma_e} \right) +
    O\left(\frac{1}{n^2}\right)
\end{align*}
\end{mylemma}
\begin{proof}
Let $B_{ij}(z) = \sign(\Delta_x) \cdot c_{ij}(z)$, where $c_{ij} = \sign (e_j - e_i - z \cdot \Delta_x)$.
Starting with the result from Lemma~\ref{lem:u-bn-statistic-variance}, we have that
\begin{align*}
    \Var_{\es}[U_z] 
    \leq \left( \frac{1}{N^2}
     \sum_{\substack{\{i, j\} \in \matchedindexset\\ \{s, t\} \in \matchedindexset\\ |\{i, j\} \cap \{s, t\}| = 1}} \Cov \left(B_{ij}(z), B_{st}(z) \right) \right) + \frac{1}{N}
\end{align*}
Note that under our additional assumptions \ref{assumption:endpoints} and \ref{assumption:normal-errors}, $\sign(\Delta_x) = 1$ and $N = \lfloor n/2 \rfloor \cdot \lceil n/2 \rceil$. 

Next, for $|\{i, j\} \cap \{s, t\}| = 1$ (and supposing $i=s$), we have that
\begin{align*}
    \Cov \left( B_{ij}(z), B_{st}(z) \right)
    &\leq \E_{\es}\left[B_{ij}(z) \cdot B_{st}(z) \right] \\
    &= \E_{e_i, e_j, e_t} \left[ c_{ij}(z) \cdot c_{it}(z) \right] \\
    &= \E_{e_i} \left[ E_{e_j, e_t} \left[ c_{ij}(z) \cdot c_{it}(z) \mid e_i \right] \right] \\
    &= \E_{e_i} \left[ E_{e_j}[c_{ij}(z) \mid e_i] \cdot E_{e_t}[c_{it}(z) \mid e_i] \right] \\
    &= \E_{e_i} \left[ (1-2 \cdot F_e(e_i + z\cdot\Delta_x))^2 \right] 
\end{align*}
The same expression follows for the case where $j = t$ instead of $i = s$. Note that $F_e$ is the cumulative distribution function for $e_i, i \in [n]$, which by Assumption~\ref{assumption:normal-errors} is $\calN(0, \sigma_e^2)$.

Setting $w = z\Delta_x$, let $g(w) = \E_{e_i} \left[ (1-2 \cdot F_e(e_i + w))^2 \right]$. Now, we can rewrite $e_i$ as $F_e^{-1}(U)$, where $U$ is a uniform $[0,1]$ random variable, and then compute $g(0)$ as follows.
\begin{align*}
    g(w) &= \E_{U} \left[ \left(1-2 \cdot F_e\left(F_e^{-1}(U) + w \right)\right)^2 \right] \\
    g(0) 
    &= E_{U}\left[ (1-2 \cdot F_e (F_e^{-1}(U)))^2 \right] 
    = E_{U}\left[ (1-2 \cdot U)^2 \right] 
    = 1/3
\end{align*}
Next, we know that for any constant $c > 0$, if $|g'(w)| \leq c$ for all $w \in \reals$, then
\begin{align*}
    g(w) \in
    \left(g(0) - cw, g(0) + cw \right)
\end{align*}
To bound $g'(w)$, we can first express it as follows.
\begin{align*}
    g'(w) 
    &= \E_{U} \left[ 2 \cdot \left(1-2 \cdot F_e\left(F_e^{-1}(U) + w\right) \right) \cdot 
    \left( -2 \cdot F_e'\left(F_e^{-1}(U) + w\right) \right) \right] \\
    &= \E_U \left[ -4 \cdot \left(1-2 \cdot F_e\left(F_e^{-1}(U) + w\right) \right) \cdot \frac{1}{\sigma_e \cdot \sqrt{2\pi}} \cdot \exp\left( \frac{-\left(F_e^{-1}(U) + w\right)^2}{2 \sigma_e^2} \right) \right] 
\end{align*}
where the second line follows by noting that $F_e'$ is a pdf of a $\calN(0,\sigma_e^2)$ random variable. 
As a cumulative distribution function, $F_e(\cdot)$ is between $0$ and $1$. In addition, we can see that the $\exp$ factor is between $0$ and $1$. Therefore, we have that
\begin{align*}
    |g'(w)| \leq 4 \cdot \frac{1}{\sigma_e \cdot \sqrt{2\pi}} = O \left(\frac{1}{\sigma_e} \right),
    ~~~\text{so}~~
    g(w) = g(0) + O \left(\frac{w}{\sigma_e} \right)
\end{align*}
Finally, the number of terms that share one datapoint, ie. $\# \{i, j\} \in \matchedindexset, \{s, t\} \in \matchedindexset$ s.t. $|\{i, j\} \cap \{s, t\}| = 1$, divided by the $\numpairs^2$ factor, is less than or equal to
\[
\frac{2 \cdot (n/2)^2 \cdot (n/2-1)}{(n/2)^4} = \frac{4(n-2)}{n^2}
\]
Putting this all together, we have that for 
$w = z \Delta_x$, the following holds.
\begin{align*}
    \Var_{\es} \left[U(\guessn, \datatuples) \right]
    &\leq \left( \frac{1}{N^2}
     \sum_{\substack{\{i, j\} \in \matchedindexset\\ \{s, t\} \in \matchedindexset\\ |\{i, j\} \cap \{s, t\}| = 1}} \Cov \left(B_{ij}(z), B_{st}(z) \right) \right) + \frac{1}{N} \\
     &\leq \left( \frac{1}{N^2}
     \sum_{\substack{\{i, j\} \in \matchedindexset\\ \{s, t\} \in \matchedindexset\\ |\{i, j\} \cap \{s, t\}| = 1}} \E_{e_i} \left[ (1-2 \cdot F_e(e_i + z\cdot\Delta_x))^2 \right] \right) + \frac{1}{N} \\
    &= \left( \frac{1}{N^2}
     \sum_{\substack{\{i, j\} \in \matchedindexset\\ \{s, t\} \in \matchedindexset\\ |\{i, j\} \cap \{s, t\}| = 1}} g(w) \right) + \frac{1}{N} \\
    &\leq \frac{1}{(n/2)^4} \cdot 2 \cdot (n/2)^2 \cdot (n/2 - 1) \cdot
     \left( \frac{1}{3} + O \left(\frac{w}{\sigma_e}\right) \right) + \frac{1}{(n/2)^2} \\
    &=
    \frac{4(n-2)}{n^2} \left( \frac{1}{3} + O\left( \frac{w}{\sigma_e} \right) \right) +
    O\left(\frac{1}{n^2}\right) \\
    &\leq
    \frac{4}{3n} + O\left( \frac{z \cdot \Delta_x}{n \cdot \sigma_e} \right) +
    O\left(\frac{1}{n^2}\right)
\end{align*}
\end{proof}

Using the above characterizations of the expectation and variance, we can now restate and prove the convergence bound for this special case.
\begin{thmn}[\ref{thm:wemts-bound-endpoints}]
Let $x_1, \ldots, x_n$ be at two endpoints of an interval of size $\Delta_x$ such that they satisfy Assumption~\ref{assumption:endpoints}, and let $e_1, \ldots, e_n$ be drawn i.i.d from $F_e = \calN(0, \sigma_e^2)$ according to Assumption~\ref{assumption:normal-errors}. Let $y_1, \ldots, y_n$ be the corresponding response variables under the model $y_i = \intercept + \slope x_i + e_i$, where $\intercept, \slope \in \reals$.
Let $\betatildewemts = \texttt{DPTheilSen}(\{x_i, y_i\}_{i=1}^n, \eps, (\wexpm, \theta, -R, R))$, where $\eps, R, \theta >0$, and $\slope \in [-R+\theta, R-\theta]$ as in Assumption~\ref{assumption:range}.
Let $\tau$ be defined as follows.
\begin{align*}
    \tau = \Phi^{-1} \left( 1 - \frac{p}{8} \right) \cdot \sqrt{\frac{4}{3n}} + \frac{8\ln (4R/p\theta)}{\eps n} 
\end{align*}
where $\Phi^{-1}$ is the inverse standard normal distribution function.
Then, we have that for sufficiently large $n$, where for some constant $c$, $n \geq c \left( \log(R/p\theta) / \eps \sqrt{\log(1/p)} \right)^2$, and sufficiently small $\tau$, we have that with probability at least $1-p$, 
\[
\betatildewemts \in [\slope - z - \theta, \slope + z + \theta]
\]
for
\begin{align*}
    z = \frac{\sqrt{\pi} \cdot \sigma_e}{\Delta_x} \cdot \left( \tau + O\left(\tau^{3/2}\right) \right).
\end{align*}
\end{thmn}
\begin{proof}
We start with the bound from Theorem~\ref{thm:wemts-bound} and substitute the expressions for $\mu(z)$ and $\sigma^2(z)$ using Lemmas~\ref{lem:u-endpoints-expectation} and~\ref{lem:u-endpoints-variance}. 
This gives the following: for sufficiently large $n$ and sufficiently small $z \cdot \Delta_x / \sigma_e$,
$\betatildewemts \in [\slope - z - \theta, \slope + z + \theta]$ provided that 
\begin{align}
\label{eq:z-condition}
    z \geq \frac{\sqrt{\pi} \cdot \sigma_e}{\Delta_x} \cdot \left( \Phi^{-1} \left( 1 - \frac{p}{4} + \frac{c_1}{n^2 \cdot \sigma^3(z)} \right) \cdot \sqrt{ \left( \frac{4}{3n} + e_{\sigma^2}(z) \right)} + \frac{8 \ln (4R/p\theta)}{\eps n} + e_\mu(z) \right)
\end{align}
where
\begin{align*} 
    e_{\sigma^2}(z) \leq \frac{c_2 \cdot \Delta_x \cdot z}{n \cdot \sigma_e} + \frac{c_3}{n^2} ~~~\text{and}~~~
    e_\mu(z) \leq \frac{c_4 \cdot \Delta_x^3 \cdot z^3}{\sigma_e^3}
\end{align*}
and where $c_1, c_2, c_3, c_4 > 0$ are constants. Now, we will show that $z$ can actually be set so as to satisfy condition (\ref{eq:z-condition}). In particular, it can be shown that setting $z$ as
\begin{align*}
    z := \frac{2 \sqrt{\pi} \cdot \sigma_e}{\Delta_x} \cdot \left( \tau + c_2 \cdot \tau^{3/2} \right) ~~~\text{for}~~~ \tau = \Phi^{-1} \left( 1 - \frac{p}{8} \right) \cdot \sqrt{\frac{4}{3n}} + \frac{8\ln (4R/p\theta)}{\eps n} 
\end{align*}
satisfies condition (\ref{eq:z-condition}).
To see this, first note that for any $z \in \reals$, $\sigma^2(z) \geq 4/3n$ by Lemma~\ref{lem:u-endpoints-variance}. Then, for $n \geq (6 c_1/p)^2$, we have that
\begin{align*}
    \Phi^{-1} \left( 1 - \frac{p}{4} + \frac{c_1}{n^2 \cdot \sigma^3(z)} \right) \leq \Phi^{-1} \left(1 - \frac{p}{8} \right).
\end{align*}
Using this, we can see that for $n \geq \max(3c_3 / 4, 8 c_2^2 (\Phi^{-1}(1-p/8))^2, (6 c_1/p)^2)$, we can bound the highest-order term on the right-hand side of (~\ref{eq:z-condition}) as follows.
\begin{align*}
    \frac{\sqrt{\pi} \cdot \sigma_e}{\Delta_x} \cdot \Phi^{-1} &\left( 1 - \frac{p}{4} + \frac{c_1}{n^2 \cdot \sigma^3(z)} \right) \cdot \sqrt{ \left( \frac{4}{3n} + e_{\sigma^2}(z) \right)} \\
    &\leq \frac{\sqrt{\pi} \cdot \sigma_e}{\Delta_x} \cdot \Phi^{-1} \left(1 - \frac{p}{8} \right) \cdot \left( \sqrt{\frac{4}{3n}} + \sqrt{\frac{2 \cdot c_2 \cdot \tau}{n}} + \sqrt{\frac{2 \cdot c_2^2 \cdot \tau^{3/2}}{n}} + \sqrt{\frac{c_3}{n^2}} \right) \\
    &\leq \frac{\sqrt{\pi} \cdot \sigma_e}{\Delta_x}  \cdot \tau \cdot \left( 1 + \sqrt{3 \cdot c_2 \cdot \tau} + \sqrt{3} \cdot c_2 \cdot \tau^4 \right) \\
    & \leq \frac{2\sqrt{\pi} \cdot \sigma_e}{\Delta_x}  \cdot \left(\tau + c_2  \cdot \tau^{3/2} \right) \\
    &= z 
\end{align*}
where the first inequality follows from the condition that $n > (6c_1/p)^2$, the second inequality follows for $n > 3 c_3/4$ and the definition of $\tau$, and the third inequality follows from $n > 8 (c_2 \cdot \Phi^{-1}(1-p/8))^2$. From this, we can see that there exists a setting of $z$ that satisfies condition (\ref{eq:z-condition}), which completes the proof.

\end{proof}

\subsection{Choosing the widening parameter, $\theta$}
\label{sec:choosing-theta}
Theorem~\ref{thm:wemts-bound-endpoints} offers insight on how to set $\theta$ in the \texttt{DPWideTS} algorithm, which was an open question raised by~\cite{alabi2022differentially}. For fixed $p, \eps, R, \sigma_e, \Delta_x$, and $n$ that satisfy the conditions stated in the Theorem,
we can set $\theta$ to minimize the bound as follows:
\begin{align*}
 \theta \approx \max \left( \frac{\sigma_e \cdot \ln (R \eps n \Delta_x / p \sigma_e)}{\eps n \Delta_x}, \frac{R}{p} \exp(-\eps n) \right)
\end{align*}
The first term in the max comes from allowing the two terms involving $\theta$ in Thm.~\ref{thm:wemts-bound-endpoints} to be approximately equal, whereas the second term in the max comes from upper bounding $O(\ln(R/p\theta)/\eps n)$ by some constant. The reason for having both terms is to account for both widespread and concentrated slopes.\footnote{Handling the case of concentrated slopes was~\cite{alabi2022differentially}'s original motivation for designing the widened exponential mechanism.}
The factor $\sigma_e / (n \Delta_x)$ in the first term corresponds to the standard deviation of the slopes computed by \texttt{DPWideTS}; when the slopes are highly concentrated, the first term becomes small.
The second term, however, is independent of $\sigma_e$ and $\Delta_x$, which allows $\theta$ to remain bounded away from $0$ and prevents a blowup in the convergence bound.
Note that $R, \eps, n$ and $p$ are known in practice; if the experimental design suggests that the slopes may be concentrated (e.g. if the x-values are located at one of two endpoints of an interval as in this special case), our analysis suggests setting $\theta$ to scale with the second term.

\section{Confidence Intervals}

In this section, we consider DP confidence intervals for the linear regression slope, $\slope$. We begin by defining the type of confidence intervals we are interested in.

\begin{mydefinition}[Confidence intervals for simple linear regression.]
Let $\bd = \datatuples$ be a dataset real-valued pairs, where $x_1, \ldots, x_n$ are fixed values that are not all equal, and $y_1, \ldots, y_n$ are the corresponding response variables under the model $y_i = \intercept + \slope x_i + e_i$, where $\intercept, \slope \in \reals$ and each $e_i$ is sampled i.i.d from a continuous, symmetric, mean-0 distribution $F_e$.
Let $I_{\reals}$ be the set of all intervals in $\reals$, and let $M_{\textrm{nonpriv}}: \dataspacereal \times \calH \rightarrow I_{\reals}$ be a deterministic mechanism that outputs an interval. For any $\alpha_1 \in (0,1)$, $M_{\textrm{nonpriv}}$ outputs a \emph{$(1-\alpha_1)$-confidence interval} for the slope $\slope$ if for all hyperparams $\in \calH$, and sufficiently large $n$, 
\begin{align*}
    \Pr_{\es \sim F_e} \left[ \slope \in M_{\textrm{nonpriv}}(\bd, \text{hyperparams}) \right] \geq 1-\alpha_1
\end{align*}
where the probability is over the randomness of the dataset $\bd$. 

Next, let $M_{\textrm{DP}}: \dataspacereal \times \posreals \times \calH \rightarrow I_{\reals}$ be a randomized $\eps$-DP mechanism that outputs an interval. For any $\alpha \in (0,1)$, $M_{\textrm{DP}}$ outputs a $(1-\alpha)$-\emph{DP confidence interval} for the slope $\slope$ if for all privacy loss parameters $\eps \in \posreals$, hyperparams $\in \calH$, and sufficiently large $n$,
\begin{align*}
    \Pr_{\substack{\es \sim F_e \\ M_{\textrm{DP}}}} \left[ \slope \in M_{\textrm{DP}}(\bd, \eps, \text{hyperparams}) \right] \geq 1-\alpha
\end{align*}
where the probability is over the randomness of both the dataset $\bd$ and the mechanism $M_{\textrm{DP}}$.
\end{mydefinition}

\subsection{Nonprivate Confidence Intervals for Theil-Sen}

Now, we will define the non-private Theil-Sen confidence interval~\citep{Sen68} and prove its coverage validity.

\begin{mydefinition}
\label{def:ts-ci-nonpriv}
Let $\matchedindexset$ be a set of $\numpairs$ unordered pairs of points, and let $U_0$ be the corresponding null U-statistic (defined in )Definition~\ref{def:u-statistic}, Equation \ref{eq:u-null-statistic}). For the corresponding set of slopes $\bs$, let $\fs$ be the empirical distribution function for $\bs$. Let $\sigma^2(0) = \Var_{\es}[U_0]$ as evaluated in Corollary~\ref{lem:u-null-statistic-variance}.
For any $\alpha_1 \in (0,1)$, let  $b$ be defined as follows: 
\begin{align*}
    b:= \frac{1}{2} \cdot \Phi^{-1} \left(1-\frac{\alpha_1}{8} \right) \cdot \sigma(0)
\end{align*}
 where $\Phi$ is the standard normal cdf. Let $\hat{\slope}_L = \fsinv(1/2 - b)$ and $\hat{\slope}_U = \fsinv(1/2 + b)$.
Then, the \emph{non-private $1-\alpha_1$-Theil-Sen confidence interval} is $[\hat{\slope}_L, \hat{\slope}_U]$.
\end{mydefinition}
The proof of this interval's validity can be found in Lemma~\ref{lem:ts-ci-nonpriv-validity}.

\begin{mylemma}
\label{lem:ts-ci-nonpriv-validity}
Let $x_1, \ldots, x_n$ satisfy Assumption~\ref{assumption:x-constants} and let $y_1, \ldots, y_n$ be the corresponding response variables under the model $y_i = \intercept + \slope x_i + e_i$, where $\intercept, \slope \in \reals$ and each $e_i$ is sampled i.i.d from a continuous, symmetric, mean-0 distribution $F_e$. Let $\matchedindexset$ be a set of $\numpairs$ unordered pairs of datapoints, and let $U_0$ be the corresponding null U-statistic defined in Definition~\ref{def:u-statistic} Equation (\ref{eq:u-null-statistic}). Let $\sigma^2(0) = \Var_{\es}[U_0]$ as evaluated in Corollary~\ref{lem:u-null-statistic-variance}.
For any $\alpha_1 \in (0,1)$, let $[\hat{\slope}_L, \hat{\slope}_U]$ be the non-private Theil-Sen confidence interval as in Definition~\ref{def:ts-ci-nonpriv}. Then, there exists a constant $c>0$ such that for all $n \geq c / \alpha_1^2$, 
$[\hat{\slope}_L, \hat{\slope}_U]$ is a $\left(1-\alpha_1 \right)$-confidence interval for the true slope $\slope$. 
\end{mylemma}
\begin{proof}
We will consider the upper limit of the interval, $\hat{\slope}_U$, for simplicity.
First, note that
\begin{align*}
    U(\hat{\slope}_U) &= 1-2\fs(\hat{\slope}_U) = 1-2\fs(\fsinv(1/2 + b)) \\
    &= 
    -2b = \Phi^{-1} (\alpha_1/8) \cdot \sigma(0)
\end{align*}
Next, we rely on the fact that if $\slope > \hat{\slope}_U $, $U(\slope) < U(\hat{\slope}_U)$. Therefore, we can use the distribution of $U(\slope)$ and its convergence to a standard normal via Theorem~\ref{thm:u-statistic-distribution-berry-esseen} to evaluate the probability that $\slope > \hat{\slope}_U$. 
\begin{align*}
    \Pr_{\bs} \left[ \slope > \hat{\slope}_U \right]
    &= \Pr_{\bs} \left[ U(\slope) < U\left( \hat{\slope}_U \right) \right] \\
    &\leq \Phi \left( \frac{ U\left( \hat{\slope}_U \right) }{\sigma(0)}\right) + O\left(\frac{1}{n^2 \cdot \sigma^3(0)}\right) \\
    &= \Phi \left( \frac{\Phi^{-1}(\alpha_1/8) \cdot \sigma(0)}{\sigma(0)} \right) + O\left(\frac{1}{n^2 \cdot \sigma^3(0)}\right) \\
    &= \alpha_1/8 + O\left(\frac{1}{n^2 \cdot \sigma^3(0)}\right) \\
    &= \alpha_1/8 + O\left(\frac{1}{\sqrt{n}} \right) \\
    &\leq \alpha_1/2
\end{align*}
where the second to last line holds follows from Corollary~\ref{lem:u-null-statistic-variance} and the last line holds for $n \geq c/\alpha_1^2$.
We can go through a similar argument for $\hat{\slope}_L$, which gives the desired result.
\end{proof}

\subsection{\texttt{DPTSCI} Algorithm}

Below, we provide pseudocode for the general algorithm for confidence intervals via \texttt{DPTS}. Note that this algorithm is similar to the \texttt{DPTS} point estimator (Alg~\ref{alg:ts}), except it replaces the call to \texttt{DPMed} with a call to \texttt{DPMedCI}, a DP algorithm for confidence intervals for the median.

\begin{algorithm}[hbtp!]
  \KwData{$\bd = \datatuples \in \dataspacereal$}
  \KwPrivacyparams{$\eps \in \nonnegreals$}
  \KwHyperparams{$\matchedindexset \in \binom{[n]}{2}$, \texttt{DPMedCI}, \text{hyperparams} $\in \calH$}

  $\slopes = \{\}$

  \For{each $\{i,j\} \in \matchedindexset$ (such that $x_i \neq x_j$)} {
    
        $s = (y_j - y_i)/(x_j - x_i)$
    
    Add $s$ to $\slopes$

  }
  Let $k = \max_{i \in [n]} \{\#j \in [n]: \{i, j\} \in \matchedindexset\}$
  
  $[\betatildets_L, \betatildets_U] = \texttt{DPMedCI}\left(\slopes, \eps/k, \text{hyperparams}\right)$ 
  
  \Return $[\betatildets_L, \betatildets_U]$
  \caption{\texttt{DPTSCI}: $\eps$-DP Algorithm}
  \label{alg:dpts-ci}
\end{algorithm}

\begin{mylemma}
\label{lem:dptsci-privacy}
Algorithm~\ref{alg:dpts-ci} is $\eps$-DP.
\end{mylemma}
\begin{proof}
The call to \texttt{DPMedCI} is $\eps/k$-DP by definition, so by simple composition and post-processing, \texttt{DPTSCI} is $\eps$-DP.
\end{proof}

\subsection{Union-bound confidence interval}

Now, we consider the coverage validity and width of the DP confidence interval when we use \texttt{DPWideCIUnion} (Alg~\ref{alg:dpwem-ci}) as the \texttt{DPMedCI} algorithm.

\begin{mytheorem}[Validity and width of \texttt{DPWideTSCIUnion}] 
\label{thm:wemts-ci-bound}
Let $x_1, \ldots, x_n$ satisfy Assumption~\ref{assumption:x-constants}, and let $y_1, \ldots, y_n$ be the corresponding response variables under the model $y_i = \intercept + \slope x_i + e_i$, where $\intercept, \slope \in \reals$ and each $e_i$ is sampled i.i.d from a continuous, symmetric, mean-0 distribution $F_e$. Let $\matchedindexset$ be a set of $\numpairs$ unordered datapoints. For $\alpha \in (0,1)$, $r_\alpha \in (0,1)$, $\eps, R, \theta >0$, and $\slope \in [-R+\theta, R-\theta]$ as in Assumption~\ref{assumption:range},
let 
$[\betatildewemts_L, \betatildewemts_U] =$ \\
$\texttt{DPTSCI}(\{x_i, y_i\}_{i=1}^n, \eps, \matchedindexset, \texttt{DPWideCIUnion}, (\alpha, r_\alpha, \theta, -R, R, \texttt{Union}=1))$. Then,
the interval $[\betatildewemts_L, \betatildewemts_U]$ is a $1-\alpha$-confidence interval for the true slope $\slope$ over the randomness in $\datatuples$ and \texttt{DPWideTSCIUnion}.

In addition, 
let $U_z$ be defined as in Definition~\ref{def:u-statistic}, Equation (\ref{eq:u-statistic-z-e}) according to set $\matchedindexset$.
Then, there exists a constant $v > 0$ such that for sufficiently large $n$, we have that with probability at least $1-\alpha$, $ \betatildewemts_U - \betatildewemts_L < 2z + 2\theta$ for all $z$ that satisfies the following.
\begin{align*}
    \frac{-\mu(z) - \frac{8\ln(2R/\alpha \theta)}{\eps \cdot n}}{\sigma(z)} \geq 2 \cdot \Phi^{-1}\left( 1 - \frac{\alpha}{2} + \frac{v}{n^2 \cdot \sigma^3(z)} \right)
\end{align*}
where $\mu(z) = \E[U_z], \sigma^2(z) = \Var[U(z)]$, and $\Phi$ is the standard normal cdf.
\end{mytheorem}
\begin{proof}
\label{proof:wemts-ci-bound}
First, let us prove the validity of the confidence interval \\
$[\betatildewemts_L, \betatildewemts_U]$. We consider the upper limit of the interval, $\betatildewemts_U$, for simplicity. 
First, we will bound the probability that the upper limit of the private interval fails ($\betatildewemts_U < \slope$) conditioned on the success of the upper limit of the non-private interval ($\slope < \hat{\slope}_U$). This analysis uses a translation between the $U$-statistic and the distribution function $\fs$, as well as the utility of the exponential mechanism (Theorem~\ref{thm:utility-of-wem}) with target quantile $q_U = 1/2 + b + c$.
\begin{align*}
    \Pr_{\bd, \wexpmci}&\left( \betatildewemts_U < \slope \wedge \slope \leq \hat{\slope}_U \right) \\
    &\leq \E_{\bd} \left[ \Pr_{\wexpmci} \left( \betatildewemts_U < \hat{\slope}_U \mid \bd \right) \right] \\
    &\leq \E_{\bd} \left[ \Pr_{\wexpmci} \left( U(\betatildewemts_U) > U(\hat{\slope}_U) \mid \bd \right) \right] \\
    &= \E_{\bd} \left[ \Pr_{\wexpmci} \left(1 - 2 \fs(\betatildewemts_U) > - 2b \mid \bd \right) \right] \\
    &= \E_{\bd} \left[ \Pr_{\wexpmci} \left( \fs(\betatildewemts_U) < 1/2 + b \mid \bd \right) \right] \\
    &= \E_{\bd} \left[ \Pr_{\wexpmci} \left( \fs(\betatildewemts_U) - q_U< -c \mid \bd \right) \right] \\
    &\leq \E_{\bd} \left[ \alpha_2 / 2 \right] \\
    &= \alpha_2/2
\end{align*}
We use this bound to evaluate the total failure probability of the private upper limit. We also use the failure probability of the upper limit of the \emph{non-private} confidence interval, $\hat{\slope}_U$, as characterized in Lemma~\ref{lem:ts-ci-nonpriv-validity}.
\begin{align*}
    \Pr_{\bd, \wexpmci} &\left[ \betatildewemts_U < \slope \right] \\
    &= \Pr_{\bd, \wexpmci}\left[ \betatildewemts_U < \slope \wedge \slope > \hat{\slope}_U \right] +  \Pr_{\bd, \wexpmci} \left[ \betatildewemts_U < \slope \wedge \slope \leq \hat{\slope}_U \right] \\
    &\leq \Pr_{\bd}\left[ \slope > \hat{\slope}_U \right] + \Pr_{\wexpmci} \left[ \betatildewemts_U < \hat{\slope}_U \right] \\
    &\leq \alpha_1/2 + \alpha_2/2 \\
    &= \alpha/2
\end{align*}
The same analysis holds for the other end of the interval, so this proves the validity of the confidence interval. Next, let us consider the width of the interval.
For simplicity, let us consider the event that $\betatildewemts_U > \slope + z + \theta$. 
The analysis is similar to the proof of Theorem~\ref{thm:wemts-bound}.
For all $z \in \reals$ and sufficiently large $n$, we have that
\begin{align*}
    \Pr_{Y, \wexpmci}&\left[ \betatildewemts_U > \slope + z + \theta \right] \\
    &\leq \Pr_{Y, \wexpmci} \left[ U(\betatildewemts_U - \theta) < -d \right] + \\
   & \Pr_{\substack{Y \\ \wexpmci}} \Big[ U(\betatildewemts_U - \theta ) \geq -d \wedge \betatildewemts_U - \theta > \slope + z \Big] \\
    &\leq \Pr_{Y, \wexpmci} \left[ U(\betatildewemts_U) < -d \right] +  \Pr_{Y} \left[ U(\slope + z) \geq -d \right] \\
    &\leq \Pr_{Y, \wexpmci} \left[ \fs(\betatildewemts_U) - q_U > d/2 - b - c \right] +  \Pr_{Y} \left[ U(\slope + z) \geq -d \right] \\
    &\leq \frac{R}{2\theta} \exp\left( - \eps \cdot (d/2 - b - c) \cdot N / 2 \right) +  \left( 1 - \Phi\left( \frac{ - d - \mu(z)}{\sigma(z)} \right) \right) + O\left(\frac{1}{n^2 \cdot \sigma^3(z) } \right)
\end{align*}
where $b, c$ are defined as in Algorithm~\ref{alg:dpwem-ci}. The first term in the last line follows from Theorem~\ref{thm:utility-of-wem} and the second term follows from Lemma~\ref{thm:u-statistic-distribution-berry-esseen}.
Setting this probability to $\leq \alpha$ and solving for $d$ and $z$ gives us the desired result.
We can run through a similar analysis for the event that $\betatildewemts_L < \slope - z - \theta$.
\end{proof}

\subsection{Tighter confidence interval}

Now, we consider the coverage validity of the tighter confidence interval, $\texttt{DPWideTSCI}$, obtained via the following subroutine.

\begin{algorithm}[hbtp!]
  \KwData{$\bs = (s_1, \ldots, s_{\numpairs})$}
  \KwInput{$\epsinput, \alphainput, \rangeinput, \graninput$}
  
  \For{$t_L \in \mathbb{N}, 1 \leq t_L \leq \numpairs/2 $}{
  
  $\gamma_{t_L}=1 - F_U \left( 1 - 2(t_L -1)/\numpairs \right) + \sum_{m = t_L}^{n} F'_U \left( 1- 2m/\numpairs \right) \cdot \frac{R}{\theta}\exp(- (m-t_L) \cdot \eps / 2 )$
  } \tcp{$F_U, F'_U$ are the CDF and PDF of the null U-statistic corresponding to $\bs$}
  $q_L = \max_{t_L \in \mathbb{N}, 1 \leq t_L < \lceil \numpairs/2 \rceil} \{t_L : \gamma_{t_L} \leq \alpha/2 \} / \numpairs$
  
  \For{$t_U \in \mathbb{N}, \numpairs/2 \leq t_U < \numpairs$}{
  
  $\gamma_{t_U} = F_U\left(1-2 t_U / \numpairs \right) + \sum_{m = 1}^{t_U} F'_U\left(1-2m/\numpairs\right) \cdot \frac{R}{\theta} \exp(- (t_U-m) \cdot \eps / 2 ) $
  }
  $q_U = \min_{t_U \in \mathbb{N}, \lceil \numpairs/2 \rceil \leq t_U < \numpairs} \{j: \gamma_{t_U} \leq \alpha/2 \} / \numpairs$
  
  \Return $q_L, q_U$
  
  \caption{\texttt{ComputeExpMechCITargets}}  \label{alg:exp-mech-compute-targets}
\end{algorithm}

To analyze this algorithm, we will first state a lemma that translates between the empirical distribution $\fs$ to the distribution of the null U-statistic, $F_U$.
\begin{mylemma}
\label{lem:rank-distribution-U}
Let $\bs = (s_1, \ldots, s_\numpairs) \in \reals^\numpairs$ be a multiset of (not necessarily independent) slopes computed via the \texttt{DPTSCI} algorithm, where each $s_j$ is drawn from a marginal distribution function that has true median $\slope$. Let $\fs$ be the empirical distribution of the slopes, let $U_0$ be the corresponding U-statistic (as defined in (\ref{eq:u-null-statistic}), and let $F_U, F'_U$ be the cdf and pdf, respectively, of $U_0\left(\slope, \datatuples\right)$. Then, for any $m \in [\numpairs]$, 
\begin{align*}
    \Pr \left[ \fs(\slope) = \frac{m}{N} \right]
    &= 
    F'_U\left(1 - \frac{2m}{\numpairs}\right) \\
    \Pr \left[ \fs(\slope) \leq \frac{m}{N} \right]
    &= 
    1 - F_U\left(1 - \frac{2m}{\numpairs}\right).
\end{align*}
\end{mylemma}
\begin{proof}
We can rewrite $\fs(\slope)$ as $1/2 - U_0(\slope)/2$, which directly gives that
\begin{align*}
    \Pr \left[ \fs(\slope) = \frac{m}{N} \right]
    &= 
    \Pr \left[ U_0(\slope) = 1 - \frac{2m}{N} \right] \\
    \Pr \left[ \fs(\slope) \leq \frac{m}{N} \right]
    &= 
    \Pr \left[ U_0(\slope) \geq  1 - \frac{2m}{N} \right].
\end{align*}
\end{proof}

Using the above lemma, we show the validity of the \texttt{DPWideTSCI} using \texttt{ComputeExpMechCITargets}. 
\begin{mytheorem}
\label{thm:wemtsci-validity-better}
Let $x_1, \ldots, x_n$ satisfy Assumption~\ref{assumption:x-constants}, and let $y_1, \ldots, y_n$ be the corresponding response variables under the model $y_i = \intercept + \slope x_i + e_i$, where $\intercept, \slope \in \reals$ and each $e_i$ is sampled i.i.d from a continuous, symmetric, mean-0 distribution $F_e$. Let $\matchedindexset$ be a set of unordered pairs of datapoints. For $\alpha \in (0,1)$, $r_\alpha = 1/2$ (as a default, since it won't be used), $\eps, R, \theta >0$, and $\slope \in [-R+\theta, R-\theta]$ as in Assumption~\ref{assumption:range},
let $[\betatildewemts_L, \betatildewemts_U] =$ 
$\texttt{DPTSCI}(\{x_i, y_i\}_{i=1}^n, \eps, \matchedindexset, \texttt{DPWideCI}, (\alpha, r_\alpha, \theta, -R, R, \texttt{Union}=0))$. Then,
the interval \\
$[\betatildewemts_L, \betatildewemts_U]$ is a $1-\alpha$-confidence interval for the true slope $\slope$ over the randomness in $\datatuples$ and \texttt{DPWideTSCI}.
\end{mytheorem}
\begin{proof}
For simplicity, we consider just the upper endpoint of the interval. Recall that for a given $t_U \in [\numpairs]$, $\tilde{U_\alpha}(\bs, t_U) = \wexpm(\bs, \eps/2k, (t_U/\numpairs, \theta, -R, R))$, and $\fs$ is the empirical distribution of the set of slopes $\bs$.
We will consider the probability that $\tilde{U_\alpha}(\bs, t_U) + \theta < \slope$ by 
splitting this event into two cases: first, when $\fs(\slope) \leq t_U/\numpairs$, and second, when $\fs(\slope) > t_U/\numpairs$.
\begin{align*}
    \Pr_{\bs, \wexpmci}& \left[ \tilde{U_\alpha}(\bs, t_U) + \theta < \slope \right] \\
    &= \sum_{m = 1}^{t_U}  
    \Pr_{\wexpmci} \left[ \tilde{U_\alpha}(\bs, t_U) + \theta < \slope \mid \fs(\slope) = \frac{m}{\numpairs} \right] \cdot \Pr_{\bs} \left[ \fs(\slope) = \frac{m}{\numpairs} \right]\\
    &+ \sum_{m = t_U+1}^{\numpairs}  
    \Pr_{\wexpmci} \left[ \tilde{U_\alpha}(\bs, t_U) + \theta < \slope \mid \fs(\slope) = \frac{m}{\numpairs} \right] \cdot 
    \Pr_{\bs} \left[ \fs(\slope) = \frac{m}{\numpairs} \right]
\end{align*}
We can simplify the first sum using Lemma~\ref{lem:rank-distribution-U} and the utility of the widened exponential mechanism, Theorem~\ref{thm:utility-of-wem}:
\begin{align*}
    \sum_{m = 1}^{m = t_U} \Pr_{\wexpmci} &\left[ \tilde{U_\alpha}(\bs, t_U) + \theta < \slope \mid \fs(\slope) = \frac{m}{\numpairs} \right] \cdot  \Pr_{\bs} \left[\fs(\slope) = m/N \right] \\
    &\leq \sum_{m = 1}^{m = t_U} \Pr_{\wexpmci} \left[ \tilde{U_\alpha}(\bs, t_U) < \slope \mid \fs(\slope) = \frac{m}{\numpairs} \right]  \cdot \Pr_{\bs} \left[\fs(\slope) = m/N \right] \\
    &= \sum_{m = 1}^{m = t_U} \Pr_{\wexpmci} \left[ \fs\left(\tilde{U_\alpha}(\bs, t_U) \right) < m/N  \right] \cdot F'_U \left( 1 - \frac{2m}{\numpairs} \right) \\
    &\leq \sum_{m = 1}^{m = t_U} \Pr_{\wexpmci} \left[ \frac{t_U}{\numpairs} -  \fs\left(\tilde{U_\alpha}(\bs, t_U) \right)  > \frac{t_U - m}{\numpairs} \right] \cdot F'_U \left( 1 - \frac{2m}{\numpairs} \right) \\
    &\leq \sum_{m = 1}^{m = t_U} \frac{R}{\theta} \exp \left( -(t_U - m) \cdot \eps/2 \right) \cdot F'_U \left( 1 - \frac{2m}{\numpairs} \right)
\end{align*}
For the second sum, we simply upper bound the first probability in the summation by $1$ and use Lemma~\ref{lem:rank-distribution-U} to characterize the distribution of $\ind_{\fs(\slope) = m/N}$.
\begin{align*}
    \sum_{m = t_U+1}^{m = \numpairs}  
    &\Pr_{\wexpmci} \left[ \tilde{U_\alpha}(\bs, t_U) + \theta < \slope \mid \fs(\slope) = \frac{m}{\numpairs} \right] \cdot \Pr_{\bs} \left[\fs(\slope) = \frac{m}{\numpairs} \right] \\
    &\leq \sum_{m = t_U + 1}^{m = \numpairs}  
     \Pr_{\bs} \left[\fs(\slope) = \frac{m}{\numpairs} \right] \\
     &\leq F_U \left( 1 - \frac{2 \cdot t_U}{\numpairs} \right)
\end{align*}
Therefore, we have that
\begin{align*}
   \Pr_{\bs, \wexpmci} \left( \tilde{U_\alpha}(\bs, t_U) + \theta < \slope \right)
   &\leq F_U \left( 1 - \frac{2 \cdot t_U}{\numpairs} \right) + \sum_{m = 1}^{t_U}  
     F'_U\left( 1 - \frac{2m}{\numpairs} \right) \cdot \frac{R}{\theta} \exp \left( -(t_U-m) \cdot \eps/2 \right) 
\end{align*}
A similar result holds for the other side of the interval. Then, validity of the interval follows directly from the selection of the target quantiles $t_L, t_U$ (via\texttt{ComputeExpMechCITargets}) such that the interval has at least $1-\alpha$ coverage.
\end{proof}

\vfill



\end{document}